\documentclass[prx,twocolumn,amsmath,amssymb,superscriptaddress]{revtex4-1}
\usepackage{graphicx}
\usepackage{amsthm,amssymb,upgreek}
\usepackage{array}
\usepackage{xcolor}
\usepackage{mathtools}
\usepackage{bm,times,enumitem}
\usepackage{hyperref,breakurl}
\hypersetup{colorlinks=true, urlcolor={blue!80!black}, linkcolor={blue!80!black}, citecolor={blue!80!black}}

\definecolor{dgreen}{rgb}{0.0, 0.5, 0.0}
\definecolor{darksienna}{rgb}{0.24, 0.08, 0.08}

\newcommand{\CORR}[1]{\textcolor{black}{#1}}

\newcommand{\ket}[1]{|{#1}\rangle}
\newcommand{\bra}[1]{\langle{#1}|}
\newcommand{\inner}[2]{\langle{#1}|{#2}\rangle}
\newcommand{\TP}[1]{{#1}^\top}
\newcommand{\rvec}[1]{\bm{#1}}
\newcommand{\dyadic}[1]{\mathbf{#1}}
\newcommand{\tr}[1]{\mathrm{tr}\{#1\}}

\begin{document}
\title{Compressively certifying quantum measurements}

\author{I. Gianani}
\affiliation{Dipartimento di Scienze, Universit{\`a} degli Studi Roma Tre, 00146 Rome, Italy}
\author{Y. S. Teo}
\email{yong.siah.teo@gmail.com}
\affiliation{Department of Physics and Astronomy, Seoul National University, 08826 Seoul, Korea}
\author{V. Cimini}
\affiliation{Dipartimento di Scienze, Universit{\`a} degli Studi Roma Tre, 00146 Rome, Italy}
\author{H. Jeong}
\affiliation{Department of Physics and Astronomy, Seoul National University, 08826 Seoul, Korea}
\author{G. Leuchs}
\affiliation{Max-Planck-Institut f\"ur die Physik des Lichts, Staudtstra\ss e 2, 91058 Erlangen, Germany}
\affiliation{Institute of Applied Physics, Russian Academy of Sciences, 603950 Nizhny Novgorod, Russia}
\author{M. Barbieri}
\affiliation{Dipartimento di Scienze, Universit{\`a} degli Studi Roma Tre, 00146 Rome, Italy}
\affiliation{Istituto Nazionale di Ottica - CNR, Largo Enrico Fermi 2, 50125 Firenze, Italy}
\author{L. L. S{\'a}nchez-Soto}
\email{lsanchez@fis.ucm.es}
\affiliation{Departamento de \'Optica, Facultad de F\'{\i}sica, Universidad Complutense, 28040 Madrid, Spain}
\affiliation{Max-Planck-Institut f\"ur die Physik des Lichts, Staudtstra\ss e 2, 91058 Erlangen, Germany}

\date{\today}

\begin{abstract}
	We introduce a reliable compressive procedure to uniquely characterize any given low-rank quantum measurement using a minimal set of probe states that is based solely on data collected from the unknown measurement itself. The procedure is most compressive when the measurement constitutes pure detection outcomes, requiring only an informationally complete number of probe states that scales linearly with the system dimension. We argue and provide numerical evidence showing that the minimal number of probe states needed is even generally below the numbers known in the closely-related classical phase-retrieval problem because of the quantum constraint. We also present affirmative results with polarization experiments that illustrate significant compressive behaviors for both two- and four-qubit detectors just by using random product probe states.
\end{abstract}
\maketitle

\section{Introduction}

Along with states and processes, measurements play a fundamental role in quantum mechanics. Physical effects observed from quantum protocols are, logically, sensitive to the actual mechanisms of the detectors~\cite{ZHANG:2013aa}, especially precision-sensitive protocols~\cite{Resch:2007a,Higgins:2007aa} and measurement-based quantum computation~\cite{NIELSEN:2003aa,Raussendorf:2001aa,Briegel:2009aa}. Unambiguous characterization of these elements is hence crucial to ensure the correct functioning of protocols in which they are employed~\cite{Luis:1999qm,Fiurasek:2001mq,D'Ariano:2004aa,D'Auria:2011aa,Zhang:2012aa,Cooper:2014qf,Altorio:2016aa,Chen:2019aa}. More precisely, a quantum measurement is modeled by a set of positive outcome operators that sum to unity, which is also known as a positive operator-valued measure~(POVM). Characterizing such a POVM entails the identification of all outcome operators by initializing input probe states and inferring these operators from the corresponding measurement data. 

\CORR{Although the explicit reconstruction of POVM outcomes has regularly been discussed for photodetectors that are described by commutative elements~\cite{Chen:2019aa,Lundeen:2009sf,Natarajan:2013bh,Schapeler:2007.16048,Bobrov:2015aa}, it cannot be overemphasized that the complete tomography of all measurement-outcome matrix representations is of paramount importance especially for general quantum information tasks that rely on non-commuting measurements. For instance, the realization of optical homodyne detectors that supply both phase and photon-number information---an interpolation between Fock states and quadrature eigenstates---requires the complete characterization of POVM outcomes to reveal their coherence properties~\cite{Zhang:2012bb,Grandi:2017aa}. As another example, in optical quantum computing where information is encoded with single-rail qubits, the complete characterization of all non-commuting projections onto the entire single-rail Bloch ball constructed from a variable displacement operator, photodetector and a feedback loop is certainly necessary~\cite{Izumi:2020aa}. In discrete-variable settings, full detector tomography has been proposed to reconstruct multiqubit POVMs with noisy off-diagonal matrix elements in near-term quantum devices as a solution to reduce read-out errors~\cite{Maciejewski:2020aa}. It should be quite obvious that partial measurements of only a few properties (like photon-loss efficiencies and cross-talk probabilities) of these highly sophisticated detectors can never be used to conclude their performance, since a huge amount of information about the off-diagonal matrix elements is lost otherwise. For this reason, the terminology ``detector tomography'' is, nowadays, practically synonymous with POVM tomography.} 
	
For $d$-dimensional quantum systems, $d^2$ probe states are necessary for this task with arbitrary measurements. However, as practical measurements of high tomographic power correspond to (nearly-)pure outcomes~\cite{Wootters:1989qf,Durt:2010cr,Scott:2006aa,Zhu:2011sp,Zhu:2014aa}, exploiting this extreme rank deficiency can significantly reduce the number of probe states. Previously, there have been proposals based on the idea of compressed sensing~\cite{Donoho:2006cs,Candes:2009cs} to reduce the measurement settings required to reconstruct low-rank quantum states~\cite{Gross:2010cs,Kalev:2015aa,Baldwin:2016cs,Riofrio:2017cs}, processes~\cite{Shabani:2011aa,Baldwin:2014aa,Rodionov:2014aa,Shabani:2011aa}, and complementary observables~\cite{Howland:2013aa,Howland:2014aa}. These proposals, nonetheless, require the correct knowledge about the maximal rank of the unknown state or process in order to choose a highly specific compressed-sensing measurement, which is difficult to justify in realistic scenarios. To the best of our knowledge, there is no compressed-sensing proposal developed for detector tomography.

\begin{figure}[t]
	\centering
	\includegraphics[width=1\columnwidth]{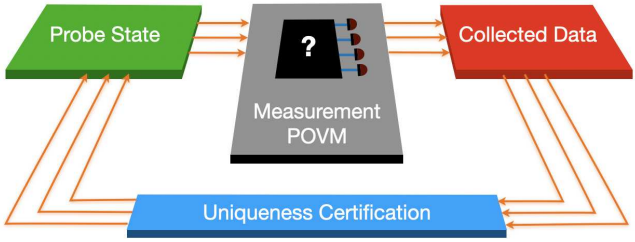}
	\caption{\label{fig:gensch}CQDT as an easy iterative procedure. A probe state is sent to the unknown POVM and its corresponding and all previous measurement data are collectively analyzed to see if they lead to a unique POVM characterization. If this is not the case, another probe state distinct from all the already chosen ones is next sent to the POVM and the procedure is repeated until a unique reconstruction is obtained.}
\end{figure}

Recently, a novel paradigm for compressive quantum state and process tomography~\cite{Ahn:2019aa,Ahn:2019ns,Teo:2020cs,Kim:2020aa} that does not depend on any spurious \emph{a priori} information has been established. It provides a built-in verification method that certifies if the characterization is truly unique from the collected data. We shall adopt a few aspects of the underlying framework to formulate the theory of compressive quantum detector tomography~(CQDT) by using a convenient set of only $O(rd)$ probe states to uniquely reconstruct any arbitrary quantum measurement of rank $r$. Interestingly, CQDT generalizes a rather extensive literature on phase-retrieval studies~\cite{ELDAR:2012aa,BANDEIRA:2014aa,Bodmann:2015aa,Xu:2018aa} where independent low-rank (positive) matrices are reconstructed from classical intensity measurements, which offers interesting mathematical results for us to benchmark our compressive scheme.

In what follows, we shall present the theory of CQDT and demonstrate its performance with several examples of low-rank quantum measurements. Furthermore, we show that the probe states needed to carry out CQDT can be very general, and the minimal number of them can even be lower than the minimal number required in phase-retrieval problems, a close cousin to the problem of CQDT, but without collective operator constraints (such as the unit-sum and positivity constraints for POVMs). In particular, we highlight that this minimal number scales linearly with $d$ for all rank-1 measurements instead of the usual quadratic behavior. To showcase CQDT in realistic physical settings, we present experimental data for both 2-qubit and 4-qubit measurements performed using polarization encoding and confirm that the resulting reconstructions are still highly compressive with real data.

\begin{figure}[t]
	\centering
	\includegraphics[width=1\columnwidth]{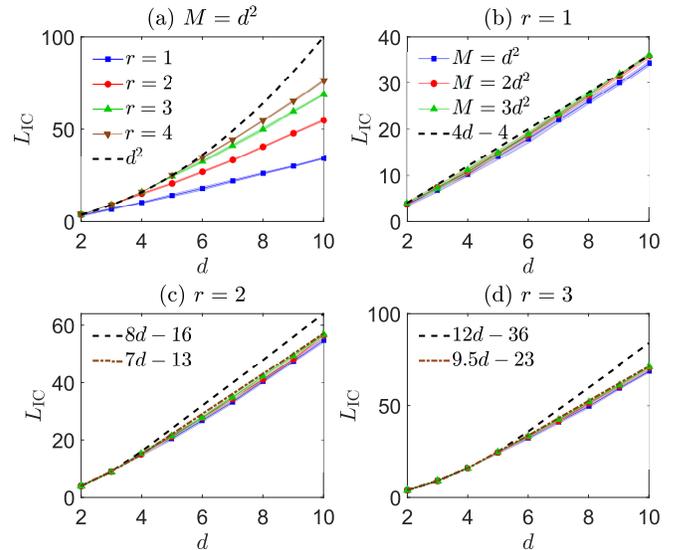}
	\caption{\label{fig:th}Plots of the IC number of rank-1 probe states $(L_\mathrm{IC})$ against dimension $(d)$ for varying values of POVM rank $r$ and $M$. (a)~CQDT on rank-1 POVMs requires only a $L_\mathrm{IC}=4d-4$ that scales linearly in $d$, whereas $\min~L$ for POVMs of higher ranks behaves as $d^2$ when $d\leq2r$ and linearly in $d$ when $d>2r$. (b,c,d)~More specifically, in comparison with the results reported in~\cite{ELDAR:2012aa,Xu:2018aa} (dashed curve) for phase retrieval, the typical number of probe states required to compressively reconstruct rank-$r$ POVMs (dotted-dashed curve) is lower as the actual POVM space is much smaller than the product of Hermitian-operator spaces when both the positivity and unit-sum constraints are simultaneously satisfied. The numerical estimates of $L_\mathrm{IC}$ pertaining to the linear regime when $d>2r$ are quoted in the legends. All graphs stabilize at the fitted functions and is verified with $M=5d^2$ (not shown in the figure panels). All 1-$\sigma$ error regions (too small to be seen here) are constructed from 10 randomly generated square-root POVMs, which are entangled measurements, and their noiseless probabilities.}
\end{figure}

\section{Compressive quantum detector tomography}

For $d$-dimensional systems, any quantum measurement, or POVM, is defined as a set of $M$ $d$-dimensional positive operators that resolve the identity $\sum^{M-1}_{j=0}\Pi_j=\openone$. Data collected with such a measurement on a given quantum state $\rho$ are statistically distributed according to the probabilities $p_j$ dictated by Born's rule: $p_j=\tr{\rho\Pi_j}$. The Hermiticity of $\Pi_j$ implies that one needs at least $d^2$ probe states to uniquely reconstruct the unknown POVM if no other additional steps are carried out.

On the other hand, common POVMs designed for quantum-information protocols are either pure or at most highly rank-deficient. To put things into perspective, for rank-$r$ operators, specified by $O(rd)\ll d^2$ parameters for $r\ll d$, it should in principle be possible to utilize $O(rd)$ probe states to uniquely characterize every single POVM element $\Pi_j$. The reconstruction is also said to be \emph{informationally complete}~(IC). The purpose of CQDT is to carry out this task without additional information about the unknown POVM (which includes its rank). It applies the uniqueness certification routine that directly inspects all data to check if a reconstruction derived from said data is unique or not.

\begin{figure}[t]
	\centering
	\includegraphics[width=1\columnwidth]{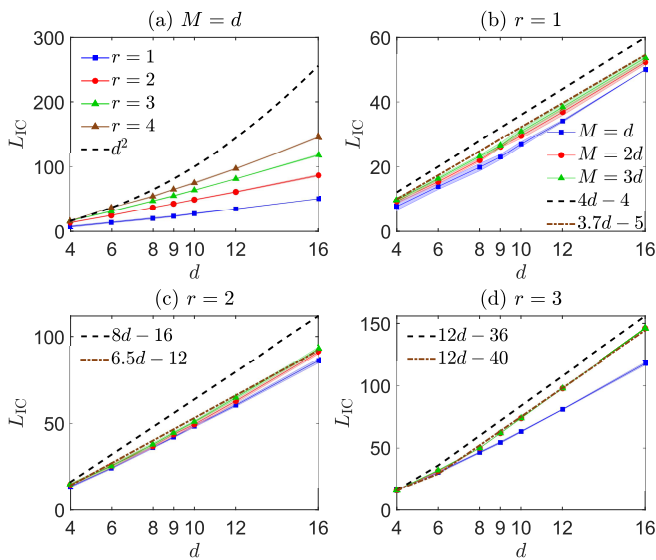}
	\caption{\label{fig:sep_th}Plots of the IC number of rank-1 product probe states $(L_\mathrm{IC})$ against dimension $(d)$ for measurement bases of various ranks $r$ and $M$. The dimensions $d$ are chosen to be either prime powers or composite numbers to showcase a variety of multipartite product probe states that are numerically investigated here. For instance, $d=4$, 8, 16 are qubit-powers, $d=9$ is a qutrit-power, and $d=6$, 10 and 12 are composite dimensions of qubits with either a qutrit or a ququint. The main figure specifications follow those of Fig.~\ref{fig:th}. The fitted expressions apply to $M=3d$, which represents a good estimate for the actual scaling behavior of $L_\mathrm{IC}$. All 1-$\sigma$ error regions (too small to be seen here) are constructed from 10 randomly generated bases and their noiseless probabilities.}
\end{figure}

\begin{figure*}[t]
	\centering
	\includegraphics[width=1.5\columnwidth]{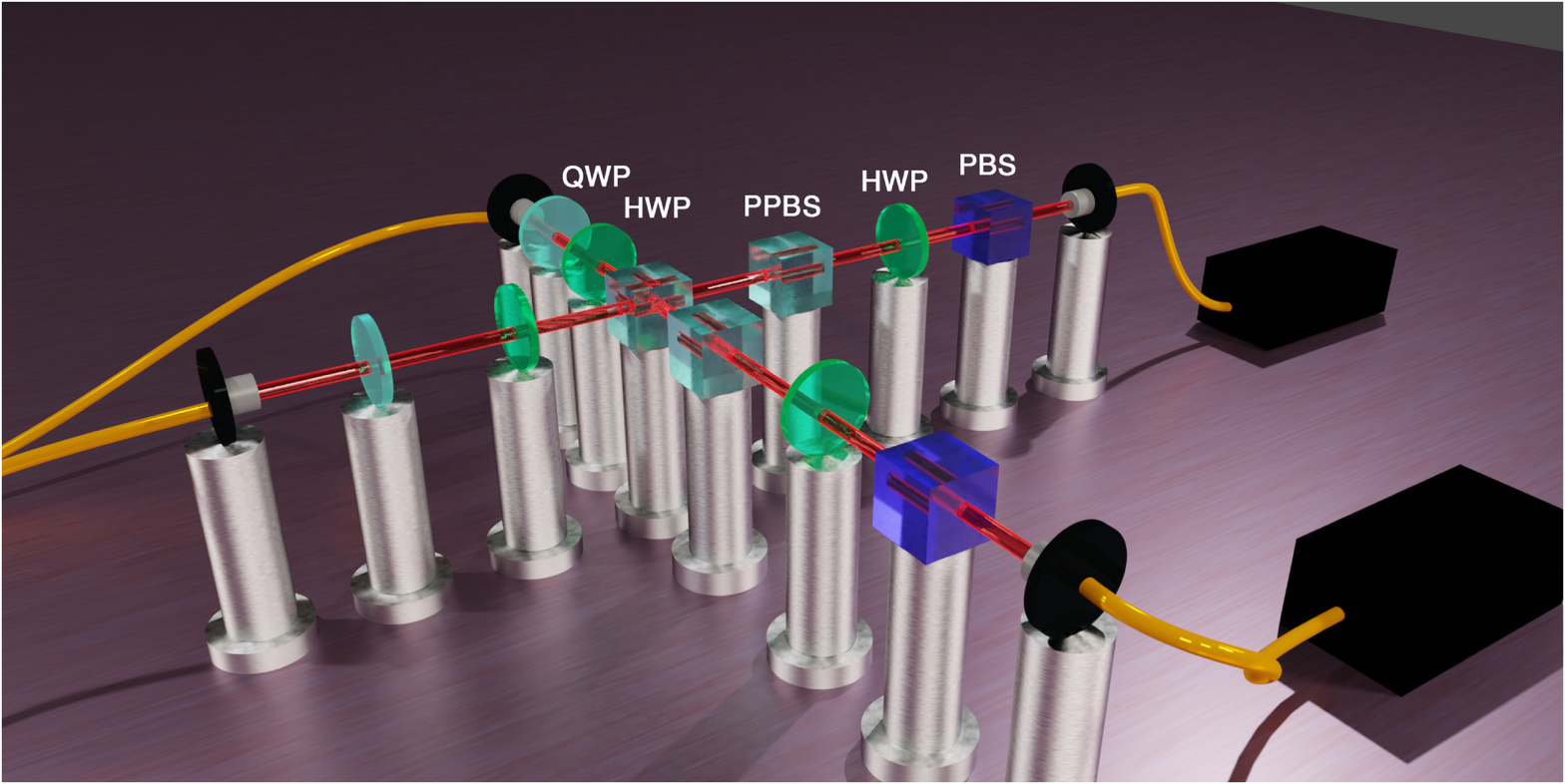}
	\caption{\label{fig:expt}Experimental setup. Photons at 810~nm are generated by SPDC from a 3~mm Type I $\beta$-barium borate crystal pumped with a 405~nm CW laser at 50~mW, on two modes selected by interference filters with FWHM = 7.3~nm and single mode fibers. Separable probe states are prepared by means of a quarter-wave plate~(QWP) at angle $\varphi_1$ followed by a half-wave plate~(HWP) at angle $\vartheta_1$ polarization rotations on one qubit in this order and a QWP at angle~$\varphi_2$ followed by a HWP at angle~$\vartheta_2$ on the other qubit in the same order. After which, the two photons are then sent through a partially polarizing beam splitter (PPBS) with transmittivities $T_H=1$ and $T_V=1/3$, acting as a controlled~Z~(C$Z$) gate. Two further PPBSs with the same transmittivities, rotated by $90^{\circ}$, are employed to compensate for the unbalance in the amplitudes of the two polarization components~\cite{Palsson}. A projective measurement is then performed on each photon by means of a HWP~($\vartheta_{m_1}$) for one output and HWP~($\vartheta_{m_2}$) for the other, and polarizing beam splitters (PBSs). The photons are then collected with single-mode fibres and sent to two avalanche photodiodes (APDs) for detection.}
\end{figure*}

To find the IC set of distinct probe states for characterizing an unknown rank-$r$ POVM in CQDT from ground up, we formulate an iterative procedure that first feeds the POVM with a randomly chosen probe state $\rho_1$. Next, the collected data $\rvec{\nu}_1=\TP{(\nu_{01}\,\,\nu_{11}\,\,\ldots\,\,\nu_{M-1\,1})}$, which are the normalized detector counts $\big($$\sum^{M-1}_{j=0}\nu_{jl}=1$ for any $l$$\big)$ distributed among the $M$ POVM outcome elements, are used to obtain the optimal physical probabilities $\widehat{\rvec{p}}_1$ that are ``nearest'' to $\rvec{\nu}_1$, where the caret denotes an estimator. This automatically defines a convex set $\mathcal{C}_1$ of POVMs that are consistent with $\widehat{\rvec{p}}_1$. The logical followup is then to verify if $\mathcal{C}_1$ has zero volume, namely whether it contains just a single POVM. Since only one probe state is used, $\mathcal{C}_1$ clearly has finite volume, so the next probe state distinct from the first is chosen and CQDT repeats, where this time the convex set $\mathcal{C}_2$ that is consistent with the probabilities $\{\widehat{\rvec{p}}_1,\widehat{\rvec{p}}_2\}$ is certified for uniqueness, and so forth (see the schematic in Fig.~\ref{fig:gensch}).

During the $L$th step of the iteration, for the sake of demonstration, we may take the optimal column of probabilities $\widehat{\rvec{p}}_l$ as the constrained least-squares~(LS) solution to the problem \begin{equation}
\min_{\left\{\Pi'_j\right\}}\left\{\sum^L_{l'=1}\|\rvec{p}_{l'}-\rvec{\nu}_{l'}\|^2\right\} 
	\quad
	\mathrm{s. \ t.} 
	\quad 
	\Pi'_j\geq0\,, \; 
	\sum^{M-1}_{j=0}\Pi'_j= \openone\,,
	\label{eq:LS}
\end{equation}
although other statistical options like maximum likelihood~\cite{Rehacek:2007ml,Teo:2011me,Shang:2017sf} may also be applied. In \eqref{eq:LS}, the $(j+1)$th entry of the column $\rvec{p}_{l'}$ is $\tr{\rho_{l'}\Pi'_j}$. After which the uniqueness certification is carried out by computing an indicator function $s_\textsc{cvx}$ over the convex set $\mathcal{C}_L$ of POVMs that are consistent with $\{\widehat{\rvec{p}}_1,\widehat{\rvec{p}}_2,\ldots,\widehat{\rvec{p}}_L\}$. A straightforward way to do this is to define $s_\textsc{cvx}=f_\text{max}-f_\text{min}$, where $f=\sum^{M-1}_{j=0}\tr{\widehat{\Pi}_jZ_j}$ and $Z_j$ are fixed but randomly-chosen full-rank positive operators. Both function optimization are carried out according to the POVM constraints and LS constraints $\big(\tr{\rho_l\widehat{\Pi}_j}=\widehat{p}_{jl}\,\,\text{and}\,\,\sum^{M-1}_{j=0}\widehat{p}_{jl}=1\,\,\text{for }0\leq j\leq M-1$ and $1\leq l\leq L\big)$. For convenience, a guided procedure is presented in Appendix~\ref{app:algo}.

Following Ref.~\cite{Ahn:2019aa}, it can be shown that if $s_\textsc{cvx}=0$, then $\mathcal{C}_L$ contains only a single unique POVM that satisfies the LS probabilities, and this is when we shall denote the IC number of probe states $L_\mathrm{IC}=L$. We note that all the physical constraints in both the LS optimization and $s_\textsc{cvx}$ computation can be conveniently integrated into semidefinite programs, which are generally polynomially efficient optimization algorithms~\cite{Vandenberghe:1996ca}.  

\begin{figure}[t]
	\centering
	\includegraphics[width=1\columnwidth]{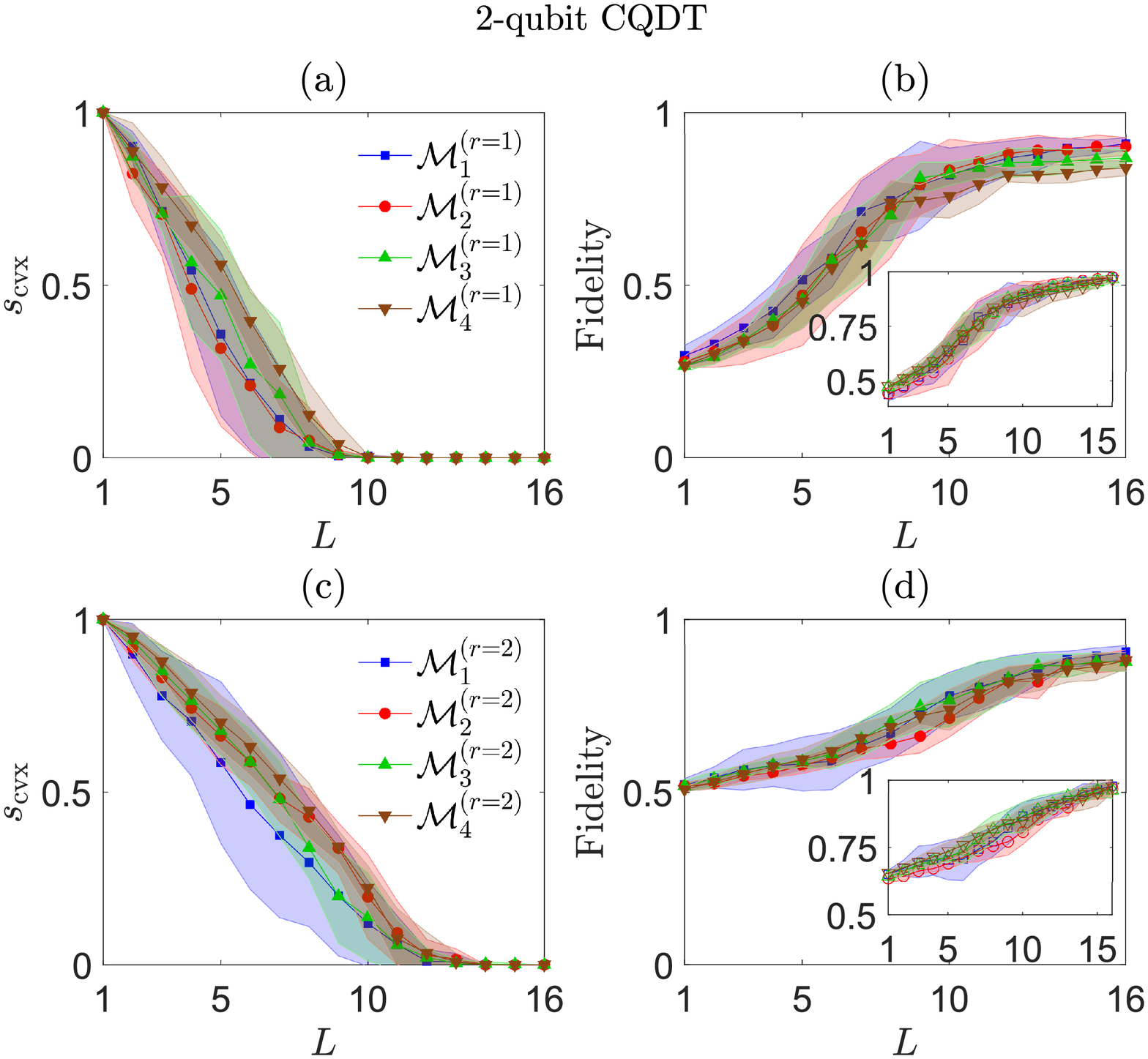}
	\caption{\label{fig:two-qubit}Two-qubit CQDT for two groups of four measurement bases $\mathcal{M}_1$ through $\mathcal{M}_4$, (a,b)~four for $r=1$ and (c,d)~another four for $r=2$. Despite the effects of statistical and systematic noise present in the experiment (the latter of which is more generally known as state-preparation-and-measurement errors~\cite{Jackson:2015aa,Ballance:2016aa,Erhard:2019aa}), the average value of $L_\mathrm{IC}$ for both $r=1$ and $r=2$, defined by the average value of $L$ at the first instance when $s_\mathrm{cvx}<10^{-3}$, closely matches the noiseless values 12 and $\approx15$ respectively. By convention, $s_\mathrm{cvx}$ is normalized by its value at $L=1$. Plots in (b) and (d) show the fidelity between the reconstructed and target POVMs, whereas their insets indicate the fidelity between the reconstructed POVM and a unique reference POVM derived from 20 probe states. All 1-$\sigma$ error regions are constructed from 10 experimental runs carried out with different probe-state sequences.}
\end{figure}

\section{Benchmarking against low-rank phase-retrieval problems} 

There is another field of study that is closely related to the problem of CQDT---the phase-retrieval problem that finds the IC set of complex signals $\{\rvec{\phi_1},\rvec{\phi_2},\ldots\}$ to uniquely identify an unknown Hermitian matrix $\dyadic{H}$ in some fixed computational basis through the respective intensity measurements $\rvec{\phi}_l^\dag\dyadic{H}\rvec{\phi}_l=y_l$~\cite{ELDAR:2012aa,BANDEIRA:2014aa,Bodmann:2015aa,Xu:2018aa}. It was conjectured in \cite{ELDAR:2012aa} and later proven in \cite{Xu:2018aa} that the IC number of signals needed to uniquely characterize a rank-$r$ $\dyadic{H}$ \emph{of known $r$} is $L^\mathrm{pr}_\mathrm{IC}=(4dr-4r^2)\eta(\lceil d/2\rceil-r)+d^2\eta(r-\lceil d/2\rceil)$ in terms of the usual Heaviside step function $\eta(\,\cdot\,)$ and the ceiling function $\lceil\,\cdot\,\rceil$ that picks the lowest integer greater than or equal to its argument.

This expression remains the same even when one attempts to recover a set of low-rank Hermitian matrices $\sum_j\dyadic{H}_j=\dyadic{1}$ that sum to the identity matrix, since this constraint merely reflects the linear dependence in the intensities $\rvec{\phi}_l^\dag\dyadic{H}_j\rvec{\phi}_l=y_{jl}$ with respect to the index $l$ and does not reduce the number of independent parameters that specify the individual matrices $\dyadic{H}_j$ except for one of them. The situation becomes starkly different when $\dyadic{H}_j\geq0$, which is that of CQDT. The positivity constraint imposed on \emph{all} matrices now heavily restricts the ranges of parameters these matrices are collectively allowed to possess in order for the unit-sum constraint to remain true. Therefore, just like quantum states and processes, compressive methods are highly effective on quantum measurements because of the positivity constraint.

\begin{figure}[t]
	\centering
	\includegraphics[width=1\columnwidth]{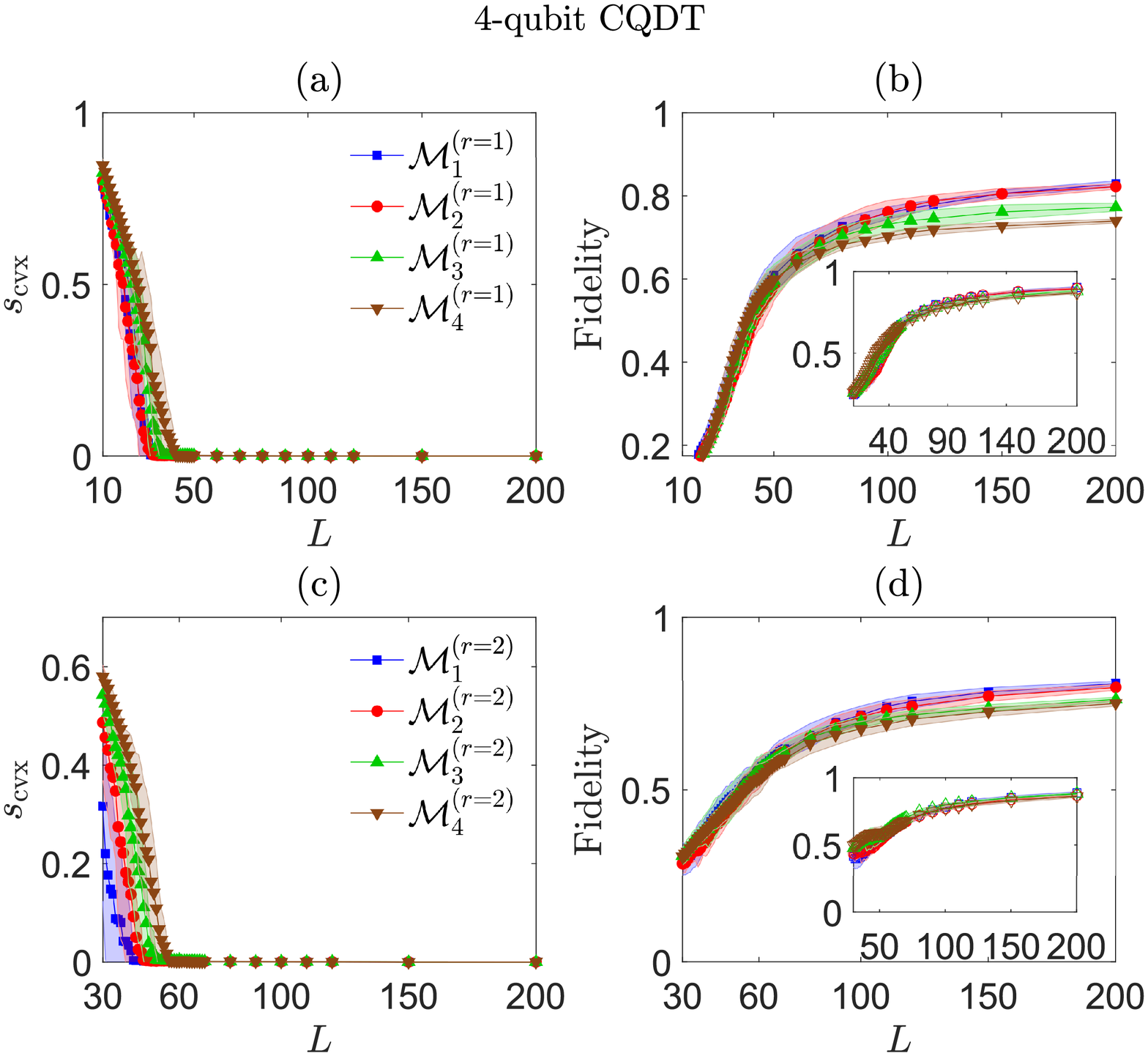}
	\caption{\label{fig:four-qubit}Four-qubit CQDT for two groups of four measurement bases $\mathcal{M}_1$ through $\mathcal{M}_4$, where the main figure specifications are identical to those of Fig.~\ref{fig:two-qubit}. Here, the higher noise levels renders the experimentally found average $L_\mathrm{IC}$ values less accurate with respect to the noiseless values for general entangled POVMs (60 for $r=1$ and $\approx99$ for $r=2$). Plots in (b) and (d) show the fidelity between the reconstructed and target POVMs, whereas their insets indicate the fidelity between the reconstructed POVM and a reference POVM that is unambiguously characterized using 256 probe states.}
\end{figure}

To gain a physical understanding of CQDT in the absence of any form of noise ($\widehat{\rvec{p}}_l=\rvec{p}_l$), Fig.~\ref{fig:th} charts the characteristic behaviors of $L_\mathrm{IC}$ with respect to the Hilbert-space dimension $d$ for low-rank POVMs. The compressive effect arising with low-rank POVMs can be observed from Fig.~\ref{fig:th}, with $L_\mathrm{IC}=4d-4=O(d)$ for rank-1 POVMs in the limit of large number ($M$) of measurement outcomes where all projectors behave approximately as independent rank-1 operators despite the unit-sum constraint. Additionally, this number is believed to be near optimal~\cite{Xu:2018aa}. In this case, $L_\mathrm{IC}\rightarrow L^\mathrm{pr}_\mathrm{IC}$ asymptotically since any rank-1 Hermitian operator $\widehat{\Pi}_j=\ket{\phi_j}\alpha_j\bra{\phi_j}$ can be written as a real-scalar multiple $(\alpha_j)$ of a projector $\ket{\phi_j}\bra{\phi_j}$, and the only difference between rank-1 phase-retrieval and CQDT is the constraint $\alpha_j>0$ for all $j$ such that enforcing this constraint does not reduce the number of parameters needed to be specified. On closer inspection of Figs.~\ref{fig:th}(b,c,d), it turns out that $L_\mathrm{IC}<L^\mathrm{pr}_\mathrm{IC}$ even in the large-$M$ limit. This time, unlike the $r=1$ case, imposing positivity on all $r$ eigenvalues of every rank-$r$ operator significantly reduces the volume of all individual linear-operator spaces. We emphasize that Fig.~\ref{fig:th} illustrates results based on randomly chosen square-root POVMs, which are ``pretty good'' measurements when employed in quantum-state discrimination problems~\cite{Hausladen:1994aa,Eldar:2001aa,Pozza:2015aa} and is interestingly equivalent to \emph{Haar-random POVMs} introduced recently in~\cite{Heinosaari:2020aa} (see also Appendix~\ref{app:SRM} for a brief recipe to generate them). The enhancement in the compressibility of CQDT as a consequence of operator constraints is a rather general quantum phenomenon~\cite{Kalev:2015aa} that manifests itself in any sort of physical measurements.

A universal feature of CQDT is that \emph{any} set of distinct probe states will serve equally well as resources for characterizing measurements. This means that product states may also be used for this purpose, not just entangled ones. This feature exists also in phase-retrieval problems (refer, for instance, to \cite{Xu:2018aa} for arguments without any explicit assumption about the intensity measurements). The underlying reason is that the degree of linear independence in the probe states has nothing to do with their entanglement content: one can find a complete set of product/separable operator basis that spans the entire linear-operator space just as well as an entangled basis. Therefore, one should expect that the scaling behavior for $L_\mathrm{IC}$ remains the same even for product probe states and a rank-$r$ subspace. This allows one to perform CQDT without entangling operations acting on the probe states. Figure~\ref{fig:sep_th} precisely confirms this intuition.

\section{Experimental confirmation}
\label{sec:expt}

We formally demonstrate CQDT using an experimental setup as shown in Fig.~\ref{fig:expt}. Two qubits are encoded in the polarization degree of freedom for photon pairs generated \emph{via} SPDC, with $\ket{\textsc{h}} \equiv \ket{1}$ and  $\ket{\textsc{v}} \equiv \ket{0}$. By means of half wave plates (HWPs) and quarter wave plates (QWPs) we prepare twenty random 2-qubit probe states as  $U_\textsc{hwp}(\vartheta_1)U_\textsc{qwp}(\varphi_1)\otimes U_\textsc{hwp}(\vartheta_2)U_\textsc{qwp}(\varphi_2)\ket{1}_1\ket{1}_2$, where the values of the waveplates angles vary in the interval  $-\pi/2\leq\vartheta_1,\vartheta_2\leq\pi/2$ and $-\pi/4\leq\varphi_1,\varphi_2\leq\pi/4$ (see Appendix \ref{app:expt} for further details).

The measurement relies on a controlled~$Z$~(C$Z$) gate, which is implemented by means of a partially polarizing beam splitter (PPBS)~\cite{nathan, kiesel, okamoto,Roccia:2017aa}, acting as $ U_\textsc{c\tiny$Z$}= \ket{0}\bra{0} \otimes\sigma_z+ \ket{1}\bra{1} \otimes \openone$ in terms of the Pauli operator $\sigma_z$. After the gate, a projective measurement is eventually performed for each qubit by means of a HWP at an angle $\vartheta_{m_1}$ for the first qubit, another HWP at an angle $\vartheta_{m_2}$ for the second, and polarizing beam splitters (PBSs). 
We consider four different POVMs, $\mathcal{M}^{(r=1)}_i=\{\Pi_j^i=\ket{\psi^i_j}\bra{\psi^i_j}\}^3_{j=0}$ for $1\leq i\leq4$, where    
\begin{equation}
\ket{\psi^i_j}=U_\textsc{c\tiny$Z$}[U_\textsc{hwp}(\vartheta_{m_1}^i)\otimes U_\textsc{hwp}(\vartheta_{m_2}^i)]\ket{l}_1\ket{l'}_2
\end{equation}
with $\ket{l\,l'}\in\{\ket{00},\ket{01},\ket{10},\ket{11}\}$, obtained by fixing the projection on the first qubit at $\vartheta_{m_1}=22.5^{\circ}$ (quoted in degrees), and adopting for the second qubit the four settings $\vartheta_{m_2}^i= 0^{\circ}, 7^{\circ}, 14^{\circ}, 22.5^{\circ}$. This amounts to vary from a separable measurement when $\vartheta_{m_2}^i= 0^{\circ}$, to an entangling one when $\vartheta_{m_2}^i= 22.5^{\circ}$. We also perform CQDT on rank-2 POVMs that are defined by linear combinations of the basis outcomes inasmuch as $\mathcal{M}^{(r=2)}_i=\{\Pi^i_{j}=(\ket{\psi^i_{j}}\bra{\psi^i_{j}}+\ket{\psi^i_{j\oplus1}}\bra{\psi^i_{j\oplus1}})/2\}^3_{j=0}$, where $\oplus$ is addition modulo 4.

The performance of CQDT in terms of the uniqueness measure $s_\textsc{cvx}$ and target POVM fidelity is demonstrated in Fig.~\ref{fig:two-qubit}. The IC number of probe states $L_\mathrm{IC}$, which is obtained at the value of $L$ for which $s_\textsc{cvx}$ first drops below some small prechosen threshold, for both ranks $r=1$ and 2 match well with the simulation values in Fig.~\ref{fig:th}. To compute the POVM fidelity, we choose to compare the POVM Choi-Jamio{\l}kowski operator~\cite{Chuang:2000fk} since the corresponding fidelity would then be invariant under arbitrary permutations of the POVM element label. For instance, the POVMs $\mathcal{M}=\{\Pi_1,\Pi_2,\Pi_3,\Pi_4\}$ and $\mathcal{M}'=\{\Pi_4,\Pi_3,\Pi_1,\Pi_2\}$ are treated as the one and the same measurement and should therefore give a unit mutual fidelity~(see Appendix~\ref{app:fid} for the technical details of the POVM fidelity computation).

To unveil how significantly compressive CQDT can get for high-dimensional systems, we also look at the performance on 4-qubit POVMs. These are derived by considering product measurements of the previous 2-qubit POVMs. Likewise, the corresponding 4-qubit input probe states are also made up of tensor-product constituents of 2-qubit probe states. The reader may consult Appendix~\ref{app:expt} for more information. The CQDT performance for these four-qubit product measurement bases are shown in Fig.~\ref{fig:four-qubit}. Owing to noise and product structures of the POVMs, we find that $L_\mathrm{IC}$ is less than the corresponding estimated values in Fig.~\ref{fig:th}. 

In both aforementioned figures, the fidelity is always less than one at $L=L_\mathrm{IC}$ because of statistical fluctuation in the data. On this note, it is instructive to recall that previous studies of overcomplete quantum tomography~\cite{Hayashi:1998aa,D'Ariano:2007aa,Bisio:2009aa,Bisio:2009qt,Zhu:2014aa,Teo:2015qs,Santis:2019aa} has led to an understanding that measuring probe states of numbers beyond $L_\mathrm{IC}$ should generally lead to an improvement in reconstruction fidelity. This is evidently observed in both Figs.~\ref{fig:two-qubit} and \ref{fig:four-qubit}.

\begin{figure}[t]
	\centering
	\includegraphics[width=1\columnwidth]{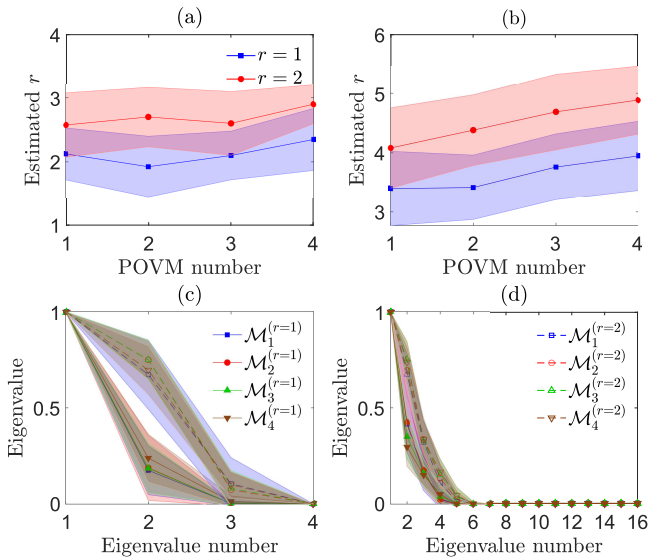}
	\caption{\label{fig:rnk_eigs}Plots of ranks and eigenvalue spectra from compressive reconstructions ($s_\textsc{cvx}<10^{-3}$) of the same groups of POVMs for both the (a,c)~2-qubit and (b,d)~4-qubit systems as in Figs.~\ref{fig:two-qubit} and \ref{fig:four-qubit}. Eigenvalue spectra of POVMs of true ranks $r=1$ (solid markers and lines) and $r=2$ (open markers and dashed lines) are shown. The 1-$\sigma$ error regions in (a) and (b) are computed over all experiments and reconstructed outcomes in each experiment, and those in (c) and (d) also include all the different POVMs.} 
\end{figure}

Figure~\ref{fig:rnk_eigs} presents plots for some more linear-algebraic properties of the reconstructed POVMs for the sake of completeness. The average rank of all reconstructed POVMs of the true rank $r=1$ is lower than that for POVMs of $r=2$, as it should be. Their deviations, however, are obvious evidence that in the presence of experimental noise, any sort of rank assumption about the unknown quantum measurement comes always with an error margin. Thus, such an assumption is never trustworthy without more extensive experimental justification. The conventional mathematical philosophy of compress-sensing that starts with a \emph{valid} rank (or its tight upper bound) of the unknown quantum measurement cannot be reliably applied in real physical situations. On the other hand, the average eigenvalue distributions for both systems indicate that all reconstructed measurement outcomes from CQDT are still fairly rank-deficient despite such deviations. This accomplishes the task of compressive characterization, and further data collection with additional probe states beyond $L=L_\text{IC}$, if the observer so chooses, would further improve the fidelity up to a subunit asymptote. To note on record, the estimated ranks for the 2-qubit ($r=1$, $r=2$) and 4-qubit ($r=1$, $r=2$) measurements, in this order, when the respective experimental datasets of $L=20$ and $L=200$ probe states are used is given by $2.250\pm0.561$, $2.875\pm0.332$, $3.644\pm0.997$ and $5.203\pm2.046$, where all statistics are computed over all POVM types, experiments and outcomes.

\section{Conclusion}

We have successfully formulated and demonstrated a highly compressive quantum detector tomography scheme that allows us to completely characterize any set of low-rank measurements using only an extremely small set of probe states relative to the square of the Hilbert-space dimension. 

To explicitly discuss its compressive performance, we have shown both numerically simulated and real polarization experimental results, which indeed confirm that data themselves permit us to know precisely whether the full measurement reconstructions are sufficiently informationally complete to unambiguously identify any given unknown quantum measurement. This works without ever depending on any kind of additional information (such as the rank) about the unknown quantum measurement, thereby making this scheme robust against noise. Furthermore, the experimental observer is free to decide whether additional probe states are necessary to further increase the target fidelity, which is unknown in practice. Hence, an approach that can aid this decision is to recognize that the fidelity ultimately saturates to a finite value that is subunity, so that the observer may choose to stop measuring more probe states once the mutual fidelity between current reconstructed measurement and the previous one approaches unity.

From the experimental results, it is also evident that product probe states can offer high compressibility for detector tomography. On hindsight, this should not come as a surprise since previous published works in (compressive) quantum tomography of various objects strongly indicate that as long as the probe states are sufficiently distinct, a unique reconstruction can still be obtained by a much smaller set of probe states.

Finally, we have emphasized the connection between quantum detector tomography and classical phase retrieval, with the former being a more general physical problem than the latter that involves additional operator constraints. Both numerical and experimental results presented here clearly show that our compressive scheme can even outperform known phase-retrieval procedures as it directly exploits the quantum positivity constraint to reduce probe-state resources. 

\begin{acknowledgments}
	We thank Emanuele Roccia for useful discussion. This work was supported in part by the National Research Foundation of Korea (NRF) (Grant Nos. NRF-2019R1A6A1A10073437, NRF-2019M3E4A1080074 and NRF-2020R1A2C1008609), the Spanish MINECO (Grant Nos. FIS201567963-P and PGC2018-099183-B-I00), and European Union's Horizon 2020 research and innovation program (Project QuantERA ApresSF). I.G. is supported by Ministero dell'Istruzione, dell'Universit{\`a} e della Ricerca Grant of Excellence Departments (ARTICOLO 1, COMMI 314-337 LEGGE 232/2016).
\end{acknowledgments}

\appendix

\section{Explicit algorithm}
\label{app:algo}

We shall state the iterative procedure of CQDT below:

\begin{center}
	\begin{minipage}[c][8cm][c]{0.9\columnwidth}
		\noindent
		\rule{\columnwidth}{1.5pt}\\
		\textbf{Compressive quantum detector tomography (CQDT)}---Starting with $l=1$ and a set of $M$ unknown POVM outcomes $\{\Pi_j\}$:
		\begin{enumerate}[series=alg]
			\item {\bf Data collection.}---Generate a probe state $\rho_l$ randomly and measure it with the unknown POVM to collect normalized data frequencies $\sum^{M-1}_{j=0}\nu_{jl}=1$ and form the column $\rvec{\nu}_l=\TP{(\nu_{0l}\,\,\nu_{1l}\,\,\ldots\,\,\nu_{M-1\,l})}$.			
			\item {\bf Physical probabilities computation.}---From the entire set of data gathered thus far ($\rvec{\nu}_1$, $\rvec{\nu}_2$, $\ldots$, $\rvec{\nu}_L$), look for their corresponding physical probabilities ($\widehat{\rvec{p}}_1$, $\widehat{\rvec{p}}_2$, $\ldots$, $\widehat{\rvec{p}}_L$), with $\widehat{\rvec{p}}_l=\TP{(\widehat{p}_{0l}\,\,\widehat{p}_{1l}\,\,\ldots\,\,\widehat{p}_{M-1\,l})}$. One may do so by solving the LS problem in \eqref{eq:LS}, or another statistical problem of choice, subject to the POVM constraints $\Pi_j\geq0$ and $\sum^{M-1}_{j=0}\Pi_j=\openone$ for $0\leq j\leq M-1$.
		\end{enumerate}
	\end{minipage}
\end{center}
\begin{center}
	\begin{minipage}[c][5cm][c]{0.9\columnwidth}
		\begin{enumerate}[resume=alg]			
			\item {\bf Uniqueness certification.}---Compute the minimum $f_\mathrm{min}$ and maximum $f_\mathrm{max}$ of the function $f=\sum^{M-1}_{j=0}\tr{\widehat{\Pi}_jZ_j}$, subject to the constraints $\Pi_j\geq0$, $\tr{\rho_l\widehat{\Pi}_j}=\widehat{p}_{jl}$ and $\sum^{M-1}_{j=0}\widehat{p}_{jl}=1$ for $0\leq j\leq M-1$ and $1\leq l\leq L$.\\[1ex]
			Define $s_\textsc{cvx}=f_\mathrm{max}-f_\mathrm{min}$. If $s_\textsc{cvx}$ is smaller than some prechosen threshold, we stop CQDT and conclude that the LS POVM is the unique estimator of the corresponding $L_\mathrm{IC}=L$. Otherwise, raise $l$ by one and repeat the whole procedure again.\\[-4ex]
		\end{enumerate}	
		\rule{\columnwidth}{1.5pt}
	\end{minipage}
\end{center}

\section{Square-root measurements}
\label{app:SRM}

There exists a simple routine to generate a POVM $\{\Pi_j\}$ whose elements $\sum_j\Pi_j=\openone$ sum to the identity. For a rank-$r$ POVM of $M$ elements:

\begin{center}
	\begin{minipage}[c][4.5cm][c]{0.9\columnwidth}
		\noindent
		\rule{\columnwidth}{1.5pt}\\
		\textbf{Square-root measurement}
		\begin{enumerate}
			\item Generate a set of $M$ operators $A_j$ represented by $d\times r$ complex matrices whose entries are independently and identically distributed according to the standard Gaussian distribution.
			\item Define $S=\sum^{M-1}_{j=0}A_jA_j^\dag$.
			\item Define $\Pi_j=S^{-1/2}A_jA_j^\dag S^{-1/2}$.\\[-5ex]
		\end{enumerate}	
		\rule{\columnwidth}{1.5pt}
	\end{minipage}
\end{center}

The above set of operators then form a POVM and is commonly coined the square-root measurement. Recently it has been shown that such measurements are in fact equivalent to Haar-random POVMs considered in~\cite{Heinosaari:2020aa}, in the sense that algebraically both kinds of measurements have identical distributions. These measurements can alternatively be generated as follows:

\begin{center}
	\begin{minipage}[c][8cm][c]{0.9\columnwidth}
		\noindent
		\rule{\columnwidth}{1.5pt}\\
		\textbf{Haar-random measurement}
		\begin{enumerate}
			\item Begin with the standard basis $\{\ket{0},\ket{1},\ldots,\ket{M-1}\}$ that spans the vector space $\mathbb{C}^M$. 
			\item Randomly sample an $rM\times d$ isometry operator $V$ $(V^\dag V=1)$ from the Haar distribution under the condition $d\leq rM$. This can be done by first generating an $rM\times rM$ complex matrix $\dyadic{A}$, then computing the QR decomposition $\dyadic{A}=\dyadic{Q}\dyadic{R}$ and defining the random Haar-distributed $rM\times rM$ unitary matrix $\dyadic{U}_\mathrm{Haar}=\dyadic{Q}\dyadic{L}$, where $\dyadic{L}=\dyadic{R}_\text{diag}\oslash|\dyadic{R}_\text{diag}|$, $\dyadic{R}_\text{diag}=\mathrm{diag}\{\dyadic{R}\}$ and $\oslash$ denotes the Hadamard division. Finally, we represent $V$ as the $rM\times d$ block of $\dyadic{U}_\mathrm{Haar}$.
			\item Define $\Pi_j=V^\dag\ket{j}\bra{j}\otimes 1_r V$ for $0\leq j\leq M-1$, where $1_r$ is the $r$-dimensional identity operator.\\[-5ex]
		\end{enumerate}	
		\rule{\columnwidth}{1.5pt}
	\end{minipage}
\end{center}

\section{Miscellaneous experimental information}
\label{app:expt}

\emph{Two-qubit state preparation.}---In Tab.~\ref{tab:multicol}, we report the wave plate settings for the preparation of the 20 random 2-qubit probe states.

\begin{table}[h!]
	\begin{center}
		\setlength{\tabcolsep}{0.5em}
		\renewcommand{\arraystretch}{2}
		\begin{tabular}{rllll}
			\hline\hline
			{\bf State}  & \quad\,\,\,$\varphi_1$ & \quad\,\,\,$\vartheta_1$ & \quad\,\,\,$\varphi_2$ & \quad\,\,\,$\vartheta_2$\\
			\hline
			{\bf1} &  $-25.95^{\circ}$ & $\hphantom{-}27.46^{\circ}$ & $-42.30^{\circ}$ & $\hphantom{-}76.53^{\circ}$ \\[-2ex]
			{\bf2} & $\hphantom{-}38.98^{\circ}$ & $-9.14^{\circ}$& $\hphantom{-}17.29^{\circ}$& $-36.51^{\circ}$\\[-2ex]
			{\bf3} & $-19.24^{\circ}$ & $\hphantom{-}20.93^{\circ}$ &  $-1.52^{\circ}$& $-60.21^{\circ}$\\[-2ex]
			{\bf4} & $-2.80^{\circ}$& $-35.81^{\circ}$& $\hphantom{-}17.10^{\circ}$& $\hphantom{-}4.65^{\circ}$\\[-2ex]
			{\bf5} & $-14.86^{\circ}$& $\hphantom{-}24.84^{\circ}$& $\hphantom{-}1.90^{\circ}$ & $-13.63^{\circ}$\\[-2ex]
			{\bf6} & $-13.00^{\circ}$& $\hphantom{-}68.55^{\circ}$ & $-5.75^{\circ}$& $-42.08^{\circ}$\\[-2ex]
			{\bf7} & $\hphantom{-}15.05^{\circ}$& $\hphantom{-}10.52^{\circ}$ & $-34.70^{\circ}$& $\hphantom{-}57.50^{\circ}$\\[-2ex]
			{\bf8} & $\hphantom{-}27.17^{\circ}$& $\hphantom{-}30.09^{\circ}$ & $-27.33^{\circ}$ & $\hphantom{-}50.72^{\circ}$\\[-2ex]
			{\bf9} & $\hphantom{-}41.99^{\circ}$& $\hphantom{-}7.78^{\circ}$& $\hphantom{-}0.73^{\circ}$& $-80.72^{\circ}$\\[-2ex]
			{\bf10} & $-18.63^{\circ}$& $\hphantom{-}28.91^{\circ}$& $\hphantom{-}0.49^{\circ}$& $\hphantom{-}20.20^{\circ}$\\[-2ex]
			{\bf11} & $-42.46^{\circ}$ & $-8.43^{\circ}$& $\hphantom{-}35.33^{\circ}$& $-58.36^{\circ}$\\[-2ex]
			{\bf12} & $-0.42^{\circ}$& $-80.64^{\circ}$ & $\hphantom{-}6.60^{\circ}$& $-79.93^{\circ}$ \\[-2ex]
			{\bf13} & $\hphantom{-}36.78^{\circ}$& $\hphantom{-}83.32^{\circ}$ & $-20.74^{\circ}$& $\hphantom{-}22.32^{\circ}$\\[-2ex]
			{\bf14} & $\hphantom{-}24.02^{\circ}$& $-55.00^{\circ}$ & $\hphantom{-}9.93^{\circ}$ & $-80.20^{\circ}$ \\[-2ex]
			{\bf15} & $\hphantom{-}21.65^{\circ}$ & $\hphantom{-}7.80^{\circ}$ & $-10.16^{\circ}$ & $\hphantom{-}6.07^{\circ}$\\[-2ex]
			{\bf16} & $-27.10^{\circ}$& $\hphantom{-}76.29^{\circ}$ & $-11.84^{\circ}$& $\hphantom{-}75.04^{\circ}$\\[-2ex]
			{\bf17} & $\hphantom{-}32.08^{\circ}$& $-39.84^{\circ}$& $-41.19^{\circ}$ & $-86.63^{\circ}$\\[-2ex]
			{\bf18} & $\hphantom{-}3.96^{\circ}$& $\hphantom{-}86.10^{\circ}$& $\hphantom{-}11.22^{\circ}$ & $-1.26^{\circ}$\\[-2ex]
			{\bf19} & $-21.22^{\circ}$& $\hphantom{-}75.01^{\circ}$ & $\hphantom{-}35.88^{\circ}$ & $\hphantom{-}68.69^{\circ}$\\[-2ex]
			{\bf20} & $\hphantom{-}14.54^{\circ}$ & $\hphantom{-}42.89^{\circ}$& $\hphantom{-}41.69^{\circ}$ & $\hphantom{-}68.41^{\circ}$\\
			\hline\hline
		\end{tabular}
	\end{center}
	\caption{{\it Experimental angular configurations (in degrees) for all optical wave plates responsible for generating the 2-qubit probe states.}}
	\label{tab:multicol}
\end{table}

\emph{Four-qubit quantum measurements.}---In the main text, we performed CQDT on 4-qubit systems by considering product and separable measurements of the 2qubit projectors described in Sec.~\ref{sec:expt}. More specifically, 4-qubit POVMs of ranks $r=1$ and 2 are defined using the 2-qubit POVMs,
\begin{align}
\mathcal{M}^{(r=1)}_i=&\,\{\Pi^i_{jk}=\ket{\psi^i_j}\bra{\psi^i_j}\otimes\ket{\psi^i_k}\bra{\psi^i_k}\}^3_{j,k=0}\,,\nonumber\\
\mathcal{M}^{(r=2)}_i=&\,\{\Pi^i_{jk}=(\ket{\psi^i_{j}}\bra{\psi^i_{j}}\otimes\ket{\psi^i_{k}}\bra{\psi^i_{k}}\nonumber\\
&\,\,\qquad+\ket{\psi^i_{j\oplus1}}\bra{\psi^i_{j\oplus1}}\otimes\ket{\psi^i_{k\oplus1}}\bra{\psi^i_{k\oplus1}})/2\}^3_{j,k=0}\,,
\end{align}
where $\oplus$ is addition modulo 4. We note that this is possible because $\inner{\psi^i_j}{\psi^i_k}=\delta_{j,k}$.

Using the 20 2-qubit probe states $\{\rho_l\}^{20}_{l=1}$, whose configurations are listed in Tab.~\ref{tab:multicol}, one can generate a total of 400 4-qubit probe states $\{\widetilde{\rho}_l\}^{400}_{l=1}$ to choose from for characterizing all $\mathcal{M}_i$s by picking $\widetilde{\rho}_l\in\{\rho_{l'}\otimes\rho_{l''},1\leq l',l''\leq20\}$. We then shuffle these 400 probe states 10 times to set up 10 experiments, each comprising a different sequence of probe states for data collection with each of the eight four-qubit POVMs $\mathcal{M}^{(r=1,2)}_i$.

With this set of probe states and product/separable measurements, all procedures of CQDT follow the prescriptions stated in Appendix~\ref{app:algo}. For a given POVM index $i$, the corresponding 4-qubit rank-1 measurement probabilities $p^i_{jkl}=\tr{\widetilde{\rho}_l\Pi^i_{jk}}=p^i_{jl'}p^i_{kl''}$ may be expressed as the product of the individual 2-qubit probabilities with the respective 2-qubit probe-state constituents, where $\widetilde{\rho}_l\equiv\rho_{l'}\otimes\rho_{l''}$. Those of the rank-2 POVMs, $p^i_{jkl}=(p^i_{jl'}p^i_{kl''}+p^i_{j\oplus1\,l'}p^i_{k\oplus1\,l''})/2$, are simply mixtures of two separate product probabilities.

\section{Fidelity between two measurements}
\label{app:fid}

We start by defining the unique square-root operators $K_j=\sqrt{\Pi_j}$ out of the POVM elements. In the language of quantum dyanmics, these form a set of Kraus operators that collectively describe the state-reduction map for the probe state $\rho$: $\rho\mapsto K_j\rho K_j^\dag/p_j$. We may then represent the POVM as a whole by a $d^2$-dimensional (trace-normalized) Choi-Jamio{\l}kowski operator $E$ by defining the canonical basis $\{\ket{0},\ket{1},\ldots\ket{d-1}\}$ and 
\begin{equation}
E=\frac{1}{d}\sum^{d-1}_{l=0}\sum^{d-1}_{l'=0}\sum^{M-1}_{j=0}K_j\ket{l}\bra{l'} K_j^\dag\otimes\ket{l}\bra{l'}\,.
\label{eq:POVM2E}
\end{equation}
Since $\tr{E}=1$, we may define the POVM fidelity $\mathcal{F}$ of two different Choi-Jamio{\l}kowski operator $E$ and $E'$ in exactly the same way as we usually do for quantum states---by means of the function $\mathcal{F}=\tr{(E^{1/2}E'E^{1/2})^{1/2}}^2$ that is symmetric in $E$ and $E'$.

It is obvious that by construction, $\mathcal{F}$ is invariant under the ordering of measurement outcomes. This benefit is, however, accompanied by an important disclaimer. Namely, $E$ is not a one-to-one representation of any POVM. This is because Eq.~\eqref{eq:POVM2E} is a result of a unidirectional mapping $\{\Pi_j\}\mapsto E$ and in the course of this procedure, information about the individual $\Pi_j$s are lost; while $\{\Pi_j\}$ guarantees a unique $E$, a given $E$ can be obtained from infinitely many sets of Kraus operators~\cite{Chuang:2000fk}. Unlike quantum processes where the Kraus operators are just mathematical representations of the unique operator $E$, quantum measurements correspond to physically singled-out Kraus operators by construction. So, although the Choi-Jamio{\l}kowski operator is ideal for computing the fidelity between two POVMs, (C)QDT cannot be performed with this operator.


\begin{thebibliography}{69}%
\makeatletter
\providecommand \@ifxundefined [1]{%
 \@ifx{#1\undefined}
}%
\providecommand \@ifnum [1]{%
 \ifnum #1\expandafter \@firstoftwo
 \else \expandafter \@secondoftwo
 \fi
}%
\providecommand \@ifx [1]{%
 \ifx #1\expandafter \@firstoftwo
 \else \expandafter \@secondoftwo
 \fi
}%
\providecommand \natexlab [1]{#1}%
\providecommand \enquote  [1]{``#1''}%
\providecommand \bibnamefont  [1]{#1}%
\providecommand \bibfnamefont [1]{#1}%
\providecommand \citenamefont [1]{#1}%
\providecommand \href@noop [0]{\@secondoftwo}%
\providecommand \href [0]{\begingroup \@sanitize@url \@href}%
\providecommand \@href[1]{\@@startlink{#1}\@@href}%
\providecommand \@@href[1]{\endgroup#1\@@endlink}%
\providecommand \@sanitize@url [0]{\catcode `\\12\catcode `\$12\catcode
  `\&12\catcode `\#12\catcode `\^12\catcode `\_12\catcode `\%12\relax}%
\providecommand \@@startlink[1]{}%
\providecommand \@@endlink[0]{}%
\providecommand \url  [0]{\begingroup\@sanitize@url \@url }%
\providecommand \@url [1]{\endgroup\@href {#1}{\urlprefix }}%
\providecommand \urlprefix  [0]{URL }%
\providecommand \Eprint [0]{\href }%
\providecommand \doibase [0]{http://dx.doi.org/}%
\providecommand \selectlanguage [0]{\@gobble}%
\providecommand \bibinfo  [0]{\@secondoftwo}%
\providecommand \bibfield  [0]{\@secondoftwo}%
\providecommand \translation [1]{[#1]}%
\providecommand \BibitemOpen [0]{}%
\providecommand \bibitemStop [0]{}%
\providecommand \bibitemNoStop [0]{.\EOS\space}%
\providecommand \EOS [0]{\spacefactor3000\relax}%
\providecommand \BibitemShut  [1]{\csname bibitem#1\endcsname}%
\let\auto@bib@innerbib\@empty
\bibitem [{\citenamefont {Zhang}\ \emph {et~al.}(2013)\citenamefont {Zhang},
  \citenamefont {Coldenstrodt-Ronge}, \citenamefont {Datta},\ and\
  \citenamefont {Walmsley}}]{ZHANG:2013aa}%
  \BibitemOpen
  \bibfield  {author} {\bibinfo {author} {\bibfnamefont {L.}~\bibnamefont
  {Zhang}}, \bibinfo {author} {\bibfnamefont {H.}~\bibnamefont
  {Coldenstrodt-Ronge}}, \bibinfo {author} {\bibfnamefont {A.}~\bibnamefont
  {Datta}}, \ and\ \bibinfo {author} {\bibfnamefont {I.~A.}\ \bibnamefont
  {Walmsley}},\ }in\ \href {https://doi.org/10.1016/B978-0-12-387695-9.00009-3}
  {\emph {\bibinfo {booktitle} {Single-Photon Generation and Detection}}},\
  \bibinfo {series} {Experimental Methods in the Physical Sciences},
  Vol.~\bibinfo {volume} {45},\ \bibinfo {editor} {edited by\ \bibinfo {editor}
  {\bibfnamefont {A.}~\bibnamefont {Migdall}}, \bibinfo {editor} {\bibfnamefont
  {S.~V.}\ \bibnamefont {Polyakov}}, \bibinfo {editor} {\bibfnamefont
  {J.}~\bibnamefont {Fan}}, \ and\ \bibinfo {editor} {\bibfnamefont {J.~C.}\
  \bibnamefont {Bienfang}}}\ (\bibinfo  {publisher} {Academic Press},\ \bibinfo
  {year} {2013})\ pp.\ \bibinfo {pages} {283 -- 313}\BibitemShut {NoStop}%
\bibitem [{\citenamefont {Resch}\ \emph {et~al.}(2007)\citenamefont {Resch},
  \citenamefont {Pregnell}, \citenamefont {Prevedel}, \citenamefont
  {Gilchrist}, \citenamefont {Pryde}, \citenamefont {O'Brien},\ and\
  \citenamefont {White}}]{Resch:2007a}%
  \BibitemOpen
  \bibfield  {author} {\bibinfo {author} {\bibfnamefont {K.~J.}\ \bibnamefont
  {Resch}}, \bibinfo {author} {\bibfnamefont {K.~L.}\ \bibnamefont {Pregnell}},
  \bibinfo {author} {\bibfnamefont {R.}~\bibnamefont {Prevedel}}, \bibinfo
  {author} {\bibfnamefont {A.}~\bibnamefont {Gilchrist}}, \bibinfo {author}
  {\bibfnamefont {G.~J.}\ \bibnamefont {Pryde}}, \bibinfo {author}
  {\bibfnamefont {J.~L.}\ \bibnamefont {O'Brien}}, \ and\ \bibinfo {author}
  {\bibfnamefont {A.~G.}\ \bibnamefont {White}},\ }\href {\doibase
  10.1103/PhysRevLett.98.223601} {\bibfield  {journal} {\bibinfo  {journal}
  {Phys. Rev. Lett.}\ }\textbf {\bibinfo {volume} {98}},\ \bibinfo {pages}
  {223601} (\bibinfo {year} {2007})}\BibitemShut {NoStop}%
\bibitem [{\citenamefont {Higgins}\ \emph {et~al.}(2007)\citenamefont
  {Higgins}, \citenamefont {Berry}, \citenamefont {Bartlett}, \citenamefont
  {Wiseman},\ and\ \citenamefont {Pryde}}]{Higgins:2007aa}%
  \BibitemOpen
  \bibfield  {author} {\bibinfo {author} {\bibfnamefont {B.~L.}\ \bibnamefont
  {Higgins}}, \bibinfo {author} {\bibfnamefont {D.~W.}\ \bibnamefont {Berry}},
  \bibinfo {author} {\bibfnamefont {S.~D.}\ \bibnamefont {Bartlett}}, \bibinfo
  {author} {\bibfnamefont {H.~M.}\ \bibnamefont {Wiseman}}, \ and\ \bibinfo
  {author} {\bibfnamefont {G.~J.}\ \bibnamefont {Pryde}},\ }\href {\doibase
  https://doi.org/10.1038/nature06257} {\bibfield  {journal} {\bibinfo
  {journal} {Nature}\ }\textbf {\bibinfo {volume} {450}},\ \bibinfo {pages}
  {393 – 396} (\bibinfo {year} {2007})}\BibitemShut {NoStop}%
\bibitem [{\citenamefont {Nielsen}(2003)}]{NIELSEN:2003aa}%
  \BibitemOpen
  \bibfield  {author} {\bibinfo {author} {\bibfnamefont {M.~A.}\ \bibnamefont
  {Nielsen}},\ }\href {\doibase https://doi.org/10.1016/S0375-9601(02)01803-0}
  {\bibfield  {journal} {\bibinfo  {journal} {Physics Letters A}\ }\textbf
  {\bibinfo {volume} {308}},\ \bibinfo {pages} {96 } (\bibinfo {year}
  {2003})}\BibitemShut {NoStop}%
\bibitem [{\citenamefont {Raussendorf}\ and\ \citenamefont
  {Briegel}(2001)}]{Raussendorf:2001aa}%
  \BibitemOpen
  \bibfield  {author} {\bibinfo {author} {\bibfnamefont {R.}~\bibnamefont
  {Raussendorf}}\ and\ \bibinfo {author} {\bibfnamefont {H.~J.}\ \bibnamefont
  {Briegel}},\ }\href {\doibase 10.1103/PhysRevLett.86.5188} {\bibfield
  {journal} {\bibinfo  {journal} {Phys. Rev. Lett.}\ }\textbf {\bibinfo
  {volume} {86}},\ \bibinfo {pages} {5188} (\bibinfo {year}
  {2001})}\BibitemShut {NoStop}%
\bibitem [{\citenamefont {Briegel}\ \emph {et~al.}(2009)\citenamefont
  {Briegel}, \citenamefont {Browne}, \citenamefont {D{\"u}r}, \citenamefont
  {Raussendorf},\ and\ \citenamefont {den Nest}}]{Briegel:2009aa}%
  \BibitemOpen
  \bibfield  {author} {\bibinfo {author} {\bibfnamefont {H.~J.}\ \bibnamefont
  {Briegel}}, \bibinfo {author} {\bibfnamefont {D.~E.}\ \bibnamefont {Browne}},
  \bibinfo {author} {\bibfnamefont {W.}~\bibnamefont {D{\"u}r}}, \bibinfo
  {author} {\bibfnamefont {R.}~\bibnamefont {Raussendorf}}, \ and\ \bibinfo
  {author} {\bibfnamefont {M.~V.}\ \bibnamefont {den Nest}},\ }\href {\doibase
  https://doi.org/10.1038/nphys1157} {\bibfield  {journal} {\bibinfo  {journal}
  {Nat. Phys.}\ }\textbf {\bibinfo {volume} {5}},\ \bibinfo {pages} {19 – 26}
  (\bibinfo {year} {2009})}\BibitemShut {NoStop}%
\bibitem [{\citenamefont {Luis}\ and\ \citenamefont
  {S\'anchez-Soto}(1999)}]{Luis:1999qm}%
  \BibitemOpen
  \bibfield  {author} {\bibinfo {author} {\bibfnamefont {A.}~\bibnamefont
  {Luis}}\ and\ \bibinfo {author} {\bibfnamefont {L.~L.}\ \bibnamefont
  {S\'anchez-Soto}},\ }\href {\doibase 10.1103/PhysRevLett.83.3573} {\bibfield
  {journal} {\bibinfo  {journal} {Phys. Rev. Lett.}\ }\textbf {\bibinfo
  {volume} {83}},\ \bibinfo {pages} {3573} (\bibinfo {year}
  {1999})}\BibitemShut {NoStop}%
\bibitem [{\citenamefont {Fiur\'a\ifmmode~\check{s}\else
  \v{s}\fi{}ek}(2001)}]{Fiurasek:2001mq}%
  \BibitemOpen
  \bibfield  {author} {\bibinfo {author} {\bibfnamefont {J.}~\bibnamefont
  {Fiur\'a\ifmmode~\check{s}\else \v{s}\fi{}ek}},\ }\href {\doibase
  10.1103/PhysRevA.64.024102} {\bibfield  {journal} {\bibinfo  {journal} {Phys.
  Rev. A}\ }\textbf {\bibinfo {volume} {64}},\ \bibinfo {pages} {024102}
  (\bibinfo {year} {2001})}\BibitemShut {NoStop}%
\bibitem [{\citenamefont {D'Ariano}\ \emph {et~al.}(2004)\citenamefont
  {D'Ariano}, \citenamefont {Maccone},\ and\ \citenamefont
  {Presti}}]{D'Ariano:2004aa}%
  \BibitemOpen
  \bibfield  {author} {\bibinfo {author} {\bibfnamefont {G.~M.}\ \bibnamefont
  {D'Ariano}}, \bibinfo {author} {\bibfnamefont {L.}~\bibnamefont {Maccone}}, \
  and\ \bibinfo {author} {\bibfnamefont {P.~L.}\ \bibnamefont {Presti}},\
  }\href {\doibase 10.1103/PhysRevLett.93.250407} {\bibfield  {journal}
  {\bibinfo  {journal} {Phys. Rev. Lett.}\ }\textbf {\bibinfo {volume} {93}},\
  \bibinfo {pages} {250407} (\bibinfo {year} {2004})}\BibitemShut {NoStop}%
\bibitem [{\citenamefont {D'Auria}\ \emph {et~al.}(2011)\citenamefont
  {D'Auria}, \citenamefont {Lee}, \citenamefont {Amri}, \citenamefont {Fabre},\
  and\ \citenamefont {Laurat}}]{D'Auria:2011aa}%
  \BibitemOpen
  \bibfield  {author} {\bibinfo {author} {\bibfnamefont {V.}~\bibnamefont
  {D'Auria}}, \bibinfo {author} {\bibfnamefont {N.}~\bibnamefont {Lee}},
  \bibinfo {author} {\bibfnamefont {T.}~\bibnamefont {Amri}}, \bibinfo {author}
  {\bibfnamefont {C.}~\bibnamefont {Fabre}}, \ and\ \bibinfo {author}
  {\bibfnamefont {J.}~\bibnamefont {Laurat}},\ }\href {\doibase
  10.1103/PhysRevLett.107.050504} {\bibfield  {journal} {\bibinfo  {journal}
  {Phys. Rev. Lett.}\ }\textbf {\bibinfo {volume} {107}},\ \bibinfo {pages}
  {050504} (\bibinfo {year} {2011})}\BibitemShut {NoStop}%
\bibitem [{\citenamefont {Zhang}\ \emph
  {et~al.}(2012{\natexlab{a}})\citenamefont {Zhang}, \citenamefont {Datta},
  \citenamefont {Coldenstrodt-Ronge}, \citenamefont {Jin}, \citenamefont
  {Eisert}, \citenamefont {Plenio},\ and\ \citenamefont
  {Walmsley}}]{Zhang:2012aa}%
  \BibitemOpen
  \bibfield  {author} {\bibinfo {author} {\bibfnamefont {L.}~\bibnamefont
  {Zhang}}, \bibinfo {author} {\bibfnamefont {A.}~\bibnamefont {Datta}},
  \bibinfo {author} {\bibfnamefont {H.~B.}\ \bibnamefont {Coldenstrodt-Ronge}},
  \bibinfo {author} {\bibfnamefont {X.-M.}\ \bibnamefont {Jin}}, \bibinfo
  {author} {\bibfnamefont {J.}~\bibnamefont {Eisert}}, \bibinfo {author}
  {\bibfnamefont {M.~B.}\ \bibnamefont {Plenio}}, \ and\ \bibinfo {author}
  {\bibfnamefont {I.~A.}\ \bibnamefont {Walmsley}},\ }\href {\doibase
  10.1088/1367-2630/14/11/115005} {\bibfield  {journal} {\bibinfo  {journal}
  {New J. Phys.}\ }\textbf {\bibinfo {volume} {14}},\ \bibinfo {pages} {115005}
  (\bibinfo {year} {2012}{\natexlab{a}})}\BibitemShut {NoStop}%
\bibitem [{\citenamefont {Cooper}\ \emph {et~al.}(2014)\citenamefont {Cooper},
  \citenamefont {Karpinski},\ and\ \citenamefont {Smith}}]{Cooper:2014qf}%
  \BibitemOpen
  \bibfield  {author} {\bibinfo {author} {\bibfnamefont {M.}~\bibnamefont
  {Cooper}}, \bibinfo {author} {\bibfnamefont {M.}~\bibnamefont {Karpinski}}, \
  and\ \bibinfo {author} {\bibfnamefont {B.~J.}\ \bibnamefont {Smith}},\ }\href
  {https://doi.org/10.1038/ncomms5332} {\bibfield  {journal} {\bibinfo
  {journal} {Nat. Commun.}\ }\textbf {\bibinfo {volume} {5}},\ \bibinfo {pages}
  {4332} (\bibinfo {year} {2014})}\BibitemShut {NoStop}%
\bibitem [{\citenamefont {Altorio}\ \emph {et~al.}(2016)\citenamefont
  {Altorio}, \citenamefont {Genoni}, \citenamefont {Somma},\ and\ \citenamefont
  {Barbieri}}]{Altorio:2016aa}%
  \BibitemOpen
  \bibfield  {author} {\bibinfo {author} {\bibfnamefont {M.}~\bibnamefont
  {Altorio}}, \bibinfo {author} {\bibfnamefont {M.~G.}\ \bibnamefont {Genoni}},
  \bibinfo {author} {\bibfnamefont {F.}~\bibnamefont {Somma}}, \ and\ \bibinfo
  {author} {\bibfnamefont {M.}~\bibnamefont {Barbieri}},\ }\href
  {http://link.aps.org/doi/10.1103/PhysRevLett.116.100802} {\bibfield
  {journal} {\bibinfo  {journal} {Phys. Rev. Lett.}\ }\textbf {\bibinfo
  {volume} {116}},\ \bibinfo {pages} {100802} (\bibinfo {year}
  {2016})}\BibitemShut {NoStop}%
\bibitem [{\citenamefont {Chen}\ \emph {et~al.}(2019)\citenamefont {Chen},
  \citenamefont {Farahzad}, \citenamefont {Yoo},\ and\ \citenamefont
  {Wei}}]{Chen:2019aa}%
  \BibitemOpen
  \bibfield  {author} {\bibinfo {author} {\bibfnamefont {Y.}~\bibnamefont
  {Chen}}, \bibinfo {author} {\bibfnamefont {M.}~\bibnamefont {Farahzad}},
  \bibinfo {author} {\bibfnamefont {S.}~\bibnamefont {Yoo}}, \ and\ \bibinfo
  {author} {\bibfnamefont {T.-C.}\ \bibnamefont {Wei}},\ }\href {\doibase
  10.1103/PhysRevA.100.052315} {\bibfield  {journal} {\bibinfo  {journal}
  {Phys. Rev. A}\ }\textbf {\bibinfo {volume} {100}},\ \bibinfo {pages}
  {052315} (\bibinfo {year} {2019})}\BibitemShut {NoStop}%
\bibitem [{\citenamefont {Lundeen}\ \emph {et~al.}(2009)\citenamefont
  {Lundeen}, \citenamefont {Feito}, \citenamefont {Coldenstrodt-Ronge},
  \citenamefont {Pregnell}, \citenamefont {Silberhorn}, \citenamefont {Ralph},
  \citenamefont {Eisert}, \citenamefont {Plenio},\ and\ \citenamefont
  {Walmsley}}]{Lundeen:2009sf}%
  \BibitemOpen
  \bibfield  {author} {\bibinfo {author} {\bibfnamefont {J.~S.}\ \bibnamefont
  {Lundeen}}, \bibinfo {author} {\bibfnamefont {A.}~\bibnamefont {Feito}},
  \bibinfo {author} {\bibfnamefont {H.}~\bibnamefont {Coldenstrodt-Ronge}},
  \bibinfo {author} {\bibfnamefont {K.~L.}\ \bibnamefont {Pregnell}}, \bibinfo
  {author} {\bibfnamefont {C.}~\bibnamefont {Silberhorn}}, \bibinfo {author}
  {\bibfnamefont {T.~C.}\ \bibnamefont {Ralph}}, \bibinfo {author}
  {\bibfnamefont {J.}~\bibnamefont {Eisert}}, \bibinfo {author} {\bibfnamefont
  {M.~B.}\ \bibnamefont {Plenio}}, \ and\ \bibinfo {author} {\bibfnamefont
  {I.~A.}\ \bibnamefont {Walmsley}},\ }\href
  {http://dx.doi.org/10.1038/nphys1133} {\bibfield  {journal} {\bibinfo
  {journal} {Nat. Phys.}\ }\textbf {\bibinfo {volume} {5}},\ \bibinfo {pages}
  {27} (\bibinfo {year} {2009})}\BibitemShut {NoStop}%
\bibitem [{\citenamefont {Natarajan}\ \emph {et~al.}(2013)\citenamefont
  {Natarajan}, \citenamefont {Zhang}, \citenamefont {Coldenstrodt-Ronge},
  \citenamefont {Donati}, \citenamefont {Dorenbos}, \citenamefont {Zwiller},
  \citenamefont {Walmsley},\ and\ \citenamefont {Hadfield}}]{Natarajan:2013bh}%
  \BibitemOpen
  \bibfield  {author} {\bibinfo {author} {\bibfnamefont {C.~M.}\ \bibnamefont
  {Natarajan}}, \bibinfo {author} {\bibfnamefont {L.}~\bibnamefont {Zhang}},
  \bibinfo {author} {\bibfnamefont {H.}~\bibnamefont {Coldenstrodt-Ronge}},
  \bibinfo {author} {\bibfnamefont {G.}~\bibnamefont {Donati}}, \bibinfo
  {author} {\bibfnamefont {S.~N.}\ \bibnamefont {Dorenbos}}, \bibinfo {author}
  {\bibfnamefont {V.}~\bibnamefont {Zwiller}}, \bibinfo {author} {\bibfnamefont
  {I.~A.}\ \bibnamefont {Walmsley}}, \ and\ \bibinfo {author} {\bibfnamefont
  {R.~H.}\ \bibnamefont {Hadfield}},\ }\href
  {http://www.opticsexpress.org/abstract.cfm?URI=oe-21-1-893} {\bibfield
  {journal} {\bibinfo  {journal} {Opt. Express}\ }\textbf {\bibinfo {volume}
  {21}},\ \bibinfo {pages} {893} (\bibinfo {year} {2013})}\BibitemShut
  {NoStop}%
\bibitem [{\citenamefont {Schapeler}\ \emph {et~al.}()\citenamefont
  {Schapeler}, \citenamefont {H{\"o}pker},\ and\ \citenamefont
  {Bartley}}]{Schapeler:2007.16048}%
  \BibitemOpen
  \bibfield  {author} {\bibinfo {author} {\bibfnamefont {T.}~\bibnamefont
  {Schapeler}}, \bibinfo {author} {\bibfnamefont {J.~P.}\ \bibnamefont
  {H{\"o}pker}}, \ and\ \bibinfo {author} {\bibfnamefont {T.~J.}\ \bibnamefont
  {Bartley}},\ }\href {https://arxiv.org/abs/2007.16048} {\ }\Eprint
  {http://arxiv.org/abs/arXiv:2007.16048} {arXiv:2007.16048} \BibitemShut
  {NoStop}%
\bibitem [{\citenamefont {Bobrov}\ \emph {et~al.}(2015)\citenamefont {Bobrov},
  \citenamefont {Kovlakov}, \citenamefont {Markov}, \citenamefont {Straupe},\
  and\ \citenamefont {Kulik}}]{Bobrov:2015aa}%
  \BibitemOpen
  \bibfield  {author} {\bibinfo {author} {\bibfnamefont {I.~B.}\ \bibnamefont
  {Bobrov}}, \bibinfo {author} {\bibfnamefont {E.~V.}\ \bibnamefont
  {Kovlakov}}, \bibinfo {author} {\bibfnamefont {A.~A.}\ \bibnamefont
  {Markov}}, \bibinfo {author} {\bibfnamefont {S.~S.}\ \bibnamefont {Straupe}},
  \ and\ \bibinfo {author} {\bibfnamefont {S.~P.}\ \bibnamefont {Kulik}},\
  }\href {\doibase 10.1364/OE.23.000649} {\bibfield  {journal} {\bibinfo
  {journal} {Opt. Express}\ }\textbf {\bibinfo {volume} {23}},\ \bibinfo
  {pages} {649} (\bibinfo {year} {2015})}\BibitemShut {NoStop}%
\bibitem [{\citenamefont {Zhang}\ \emph
  {et~al.}(2012{\natexlab{b}})\citenamefont {Zhang}, \citenamefont
  {Coldenstrodt-Ronge}, \citenamefont {Datta}, \citenamefont {Puentes},
  \citenamefont {Lundeen}, \citenamefont {Jin}, \citenamefont {Smith},
  \citenamefont {Plenio},\ and\ \citenamefont {Walmsley}}]{Zhang:2012bb}%
  \BibitemOpen
  \bibfield  {author} {\bibinfo {author} {\bibfnamefont {L.}~\bibnamefont
  {Zhang}}, \bibinfo {author} {\bibfnamefont {H.}~\bibnamefont
  {Coldenstrodt-Ronge}}, \bibinfo {author} {\bibfnamefont {A.}~\bibnamefont
  {Datta}}, \bibinfo {author} {\bibfnamefont {G.}~\bibnamefont {Puentes}},
  \bibinfo {author} {\bibfnamefont {J.~S.}\ \bibnamefont {Lundeen}}, \bibinfo
  {author} {\bibfnamefont {X.-M.}\ \bibnamefont {Jin}}, \bibinfo {author}
  {\bibfnamefont {B.~J.}\ \bibnamefont {Smith}}, \bibinfo {author}
  {\bibfnamefont {M.~B.}\ \bibnamefont {Plenio}}, \ and\ \bibinfo {author}
  {\bibfnamefont {I.~A.}\ \bibnamefont {Walmsley}},\ }\href
  {https://doi.org/10.1038/nphoton.2012.107} {\bibfield  {journal} {\bibinfo
  {journal} {Nat. Photon.}\ }\textbf {\bibinfo {volume} {6}},\ \bibinfo {pages}
  {364} (\bibinfo {year} {2012}{\natexlab{b}})}\BibitemShut {NoStop}%
\bibitem [{\citenamefont {Grandi}\ \emph {et~al.}(2017)\citenamefont {Grandi},
  \citenamefont {Zavatta}, \citenamefont {Bellini},\ and\ \citenamefont
  {Paris}}]{Grandi:2017aa}%
  \BibitemOpen
  \bibfield  {author} {\bibinfo {author} {\bibfnamefont {S.}~\bibnamefont
  {Grandi}}, \bibinfo {author} {\bibfnamefont {A.}~\bibnamefont {Zavatta}},
  \bibinfo {author} {\bibfnamefont {M.}~\bibnamefont {Bellini}}, \ and\
  \bibinfo {author} {\bibfnamefont {M.~G.~A.}\ \bibnamefont {Paris}},\ }\href
  {\doibase 10.1088/1367-2630/aa6f2c} {\bibfield  {journal} {\bibinfo
  {journal} {New J. Phys.}\ }\textbf {\bibinfo {volume} {19}},\ \bibinfo
  {pages} {053015} (\bibinfo {year} {2017})}\BibitemShut {NoStop}%
\bibitem [{\citenamefont {Izumi}\ \emph {et~al.}(2020)\citenamefont {Izumi},
  \citenamefont {Neergaard-Nielsen},\ and\ \citenamefont
  {Andersen}}]{Izumi:2020aa}%
  \BibitemOpen
  \bibfield  {author} {\bibinfo {author} {\bibfnamefont {S.}~\bibnamefont
  {Izumi}}, \bibinfo {author} {\bibfnamefont {J.~S.}\ \bibnamefont
  {Neergaard-Nielsen}}, \ and\ \bibinfo {author} {\bibfnamefont {U.~L.}\
  \bibnamefont {Andersen}},\ }\href {\doibase 10.1103/PhysRevLett.124.070502}
  {\bibfield  {journal} {\bibinfo  {journal} {Phys. Rev. Lett.}\ }\textbf
  {\bibinfo {volume} {124}},\ \bibinfo {pages} {070502} (\bibinfo {year}
  {2020})}\BibitemShut {NoStop}%
\bibitem [{\citenamefont {Maciejewski}\ \emph {et~al.}(2020)\citenamefont
  {Maciejewski}, \citenamefont {Zimbor{\'{a}}s},\ and\ \citenamefont
  {Oszmaniec}}]{Maciejewski:2020aa}%
  \BibitemOpen
  \bibfield  {author} {\bibinfo {author} {\bibfnamefont {F.~B.}\ \bibnamefont
  {Maciejewski}}, \bibinfo {author} {\bibfnamefont {Z.}~\bibnamefont
  {Zimbor{\'{a}}s}}, \ and\ \bibinfo {author} {\bibfnamefont {M.}~\bibnamefont
  {Oszmaniec}},\ }\href {\doibase 10.22331/q-2020-04-24-257} {\bibfield
  {journal} {\bibinfo  {journal} {{Quantum}}\ }\textbf {\bibinfo {volume}
  {4}},\ \bibinfo {pages} {257} (\bibinfo {year} {2020})}\BibitemShut {NoStop}%
\bibitem [{\citenamefont {Wootters}\ and\ \citenamefont
  {Fields}(1989)}]{Wootters:1989qf}%
  \BibitemOpen
  \bibfield  {author} {\bibinfo {author} {\bibfnamefont {W.~K.}\ \bibnamefont
  {Wootters}}\ and\ \bibinfo {author} {\bibfnamefont {B.~D.}\ \bibnamefont
  {Fields}},\ }\href {https://doi.org/10.1016/0003-4916(89)90322-9} {\bibfield
  {journal} {\bibinfo  {journal} {Ann. Phys.}\ }\textbf {\bibinfo {volume}
  {191}},\ \bibinfo {pages} {363} (\bibinfo {year} {1989})}\BibitemShut
  {NoStop}%
\bibitem [{\citenamefont {Durt}\ \emph {et~al.}(2010)\citenamefont {Durt},
  \citenamefont {Englert}, \citenamefont {Bengtsson},\ and\ \citenamefont
  {{\.Z}yczkowski}}]{Durt:2010cr}%
  \BibitemOpen
  \bibfield  {author} {\bibinfo {author} {\bibfnamefont {T.}~\bibnamefont
  {Durt}}, \bibinfo {author} {\bibfnamefont {B.-G.}\ \bibnamefont {Englert}},
  \bibinfo {author} {\bibfnamefont {I.}~\bibnamefont {Bengtsson}}, \ and\
  \bibinfo {author} {\bibfnamefont {K.}~\bibnamefont {{\.Z}yczkowski}},\ }\href
  {https://doi.org/10.1142/S0219749910006502} {\bibfield  {journal} {\bibinfo
  {journal} {Int. J. Quantum Inf.}\ }\textbf {\bibinfo {volume} {8}},\ \bibinfo
  {pages} {533} (\bibinfo {year} {2010})}\BibitemShut {NoStop}%
\bibitem [{\citenamefont {Scott}(2006)}]{Scott:2006aa}%
  \BibitemOpen
  \bibfield  {author} {\bibinfo {author} {\bibfnamefont {A.~J.}\ \bibnamefont
  {Scott}},\ }\href {\doibase 10.1088/0305-4470/39/43/009} {\bibfield
  {journal} {\bibinfo  {journal} {J. Phys. A: Math. and Gen.}\ }\textbf
  {\bibinfo {volume} {39}},\ \bibinfo {pages} {13507} (\bibinfo {year}
  {2006})}\BibitemShut {NoStop}%
\bibitem [{\citenamefont {Zhu}\ and\ \citenamefont
  {Englert}(2011)}]{Zhu:2011sp}%
  \BibitemOpen
  \bibfield  {author} {\bibinfo {author} {\bibfnamefont {H.}~\bibnamefont
  {Zhu}}\ and\ \bibinfo {author} {\bibfnamefont {B.-G.}\ \bibnamefont
  {Englert}},\ }\href {https://dx.doi.org/10.1103/PhysRevA.84.022327}
  {\bibfield  {journal} {\bibinfo  {journal} {Phys. Rev. A}\ }\textbf {\bibinfo
  {volume} {84}},\ \bibinfo {pages} {022327} (\bibinfo {year}
  {2011})}\BibitemShut {NoStop}%
\bibitem [{\citenamefont {Zhu}(2014)}]{Zhu:2014aa}%
  \BibitemOpen
  \bibfield  {author} {\bibinfo {author} {\bibfnamefont {H.}~\bibnamefont
  {Zhu}},\ }\href {http://dx.doi.org/10.1103/PhysRevA.90.012115} {\bibfield
  {journal} {\bibinfo  {journal} {Phys. Rev. A}\ }\textbf {\bibinfo {volume}
  {90}},\ \bibinfo {pages} {012115} (\bibinfo {year} {2014})}\BibitemShut
  {NoStop}%
\bibitem [{\citenamefont {Donoho}(2006)}]{Donoho:2006cs}%
  \BibitemOpen
  \bibfield  {author} {\bibinfo {author} {\bibfnamefont {D.}~\bibnamefont
  {Donoho}},\ }\href {https://dx.doi.org/10.1109/TIT.2006.871582} {\bibfield
  {journal} {\bibinfo  {journal} {IEEE Trans. Inf. Theory}\ }\textbf {\bibinfo
  {volume} {52}},\ \bibinfo {pages} {1289} (\bibinfo {year}
  {2006})}\BibitemShut {NoStop}%
\bibitem [{\citenamefont {Cand{\`e}s}\ and\ \citenamefont
  {Recht}(2009)}]{Candes:2009cs}%
  \BibitemOpen
  \bibfield  {author} {\bibinfo {author} {\bibfnamefont {E.~J.}\ \bibnamefont
  {Cand{\`e}s}}\ and\ \bibinfo {author} {\bibfnamefont {B.}~\bibnamefont
  {Recht}},\ }\href {https://dx.doi.org/10.1007/s10208-009-9045-5} {\bibfield
  {journal} {\bibinfo  {journal} {Found. Comput. Math.}\ }\textbf {\bibinfo
  {volume} {9}},\ \bibinfo {pages} {717} (\bibinfo {year} {2009})}\BibitemShut
  {NoStop}%
\bibitem [{\citenamefont {Gross}\ \emph {et~al.}(2010)\citenamefont {Gross},
  \citenamefont {Liu}, \citenamefont {Flammia}, \citenamefont {Becker},\ and\
  \citenamefont {Eisert}}]{Gross:2010cs}%
  \BibitemOpen
  \bibfield  {author} {\bibinfo {author} {\bibfnamefont {D.}~\bibnamefont
  {Gross}}, \bibinfo {author} {\bibfnamefont {Y.-K.}\ \bibnamefont {Liu}},
  \bibinfo {author} {\bibfnamefont {S.~T.}\ \bibnamefont {Flammia}}, \bibinfo
  {author} {\bibfnamefont {S.}~\bibnamefont {Becker}}, \ and\ \bibinfo {author}
  {\bibfnamefont {J.}~\bibnamefont {Eisert}},\ }\href
  {https://dx.doi.org/10.1103/PhysRevLett.105.150401} {\bibfield  {journal}
  {\bibinfo  {journal} {Phys. Rev. Lett.}\ }\textbf {\bibinfo {volume} {105}},\
  \bibinfo {pages} {150401} (\bibinfo {year} {2010})}\BibitemShut {NoStop}%
\bibitem [{\citenamefont {Kalev}\ \emph {et~al.}(2015)\citenamefont {Kalev},
  \citenamefont {Kosut},\ and\ \citenamefont {Deutsch}}]{Kalev:2015aa}%
  \BibitemOpen
  \bibfield  {author} {\bibinfo {author} {\bibfnamefont {A.}~\bibnamefont
  {Kalev}}, \bibinfo {author} {\bibfnamefont {R.~L.}\ \bibnamefont {Kosut}}, \
  and\ \bibinfo {author} {\bibfnamefont {I.~H.}\ \bibnamefont {Deutsch}},\
  }\href {http://dx.doi.org/10.1038/npjqi.2015.18} {\bibfield  {journal}
  {\bibinfo  {journal} {npj Quantum Inf.}\ }\textbf {\bibinfo {volume} {1}},\
  \bibinfo {pages} {15018} (\bibinfo {year} {2015})}\BibitemShut {NoStop}%
\bibitem [{\citenamefont {Baldwin}\ \emph {et~al.}(2016)\citenamefont
  {Baldwin}, \citenamefont {Deutsch},\ and\ \citenamefont
  {Kalev}}]{Baldwin:2016cs}%
  \BibitemOpen
  \bibfield  {author} {\bibinfo {author} {\bibfnamefont {C.~H.}\ \bibnamefont
  {Baldwin}}, \bibinfo {author} {\bibfnamefont {I.~H.}\ \bibnamefont
  {Deutsch}}, \ and\ \bibinfo {author} {\bibfnamefont {A.}~\bibnamefont
  {Kalev}},\ }\href {http://dx.doi.org/10.1103/PhysRevA.93.052105} {\bibfield
  {journal} {\bibinfo  {journal} {Phys. Rev. A}\ }\textbf {\bibinfo {volume}
  {93}},\ \bibinfo {pages} {052105} (\bibinfo {year} {2016})}\BibitemShut
  {NoStop}%
\bibitem [{\citenamefont {Riofr{\'i}o}\ \emph {et~al.}(2017)\citenamefont
  {Riofr{\'i}o}, \citenamefont {Gross}, \citenamefont {Flammia}, \citenamefont
  {Monz}, \citenamefont {Nigg}, \citenamefont {Blatt},\ and\ \citenamefont
  {Eisert}}]{Riofrio:2017cs}%
  \BibitemOpen
  \bibfield  {author} {\bibinfo {author} {\bibfnamefont {C.~A.}\ \bibnamefont
  {Riofr{\'i}o}}, \bibinfo {author} {\bibfnamefont {D.}~\bibnamefont {Gross}},
  \bibinfo {author} {\bibfnamefont {S.~T.}\ \bibnamefont {Flammia}}, \bibinfo
  {author} {\bibfnamefont {T.}~\bibnamefont {Monz}}, \bibinfo {author}
  {\bibfnamefont {D.}~\bibnamefont {Nigg}}, \bibinfo {author} {\bibfnamefont
  {R.}~\bibnamefont {Blatt}}, \ and\ \bibinfo {author} {\bibfnamefont
  {J.}~\bibnamefont {Eisert}},\ }\href {https://dx.doi.org/10.1038/ncomms15305}
  {\bibfield  {journal} {\bibinfo  {journal} {Nat. Commun.}\ }\textbf {\bibinfo
  {volume} {8}},\ \bibinfo {pages} {15305} (\bibinfo {year}
  {2017})}\BibitemShut {NoStop}%
\bibitem [{\citenamefont {Shabani}\ \emph {et~al.}(2011)\citenamefont
  {Shabani}, \citenamefont {Kosut}, \citenamefont {Mohseni}, \citenamefont
  {Rabitz}, \citenamefont {Broome}, \citenamefont {Almeida}, \citenamefont
  {Fedrizzi},\ and\ \citenamefont {White}}]{Shabani:2011aa}%
  \BibitemOpen
  \bibfield  {author} {\bibinfo {author} {\bibfnamefont {A.}~\bibnamefont
  {Shabani}}, \bibinfo {author} {\bibfnamefont {R.~L.}\ \bibnamefont {Kosut}},
  \bibinfo {author} {\bibfnamefont {M.}~\bibnamefont {Mohseni}}, \bibinfo
  {author} {\bibfnamefont {H.}~\bibnamefont {Rabitz}}, \bibinfo {author}
  {\bibfnamefont {M.~A.}\ \bibnamefont {Broome}}, \bibinfo {author}
  {\bibfnamefont {M.~P.}\ \bibnamefont {Almeida}}, \bibinfo {author}
  {\bibfnamefont {A.}~\bibnamefont {Fedrizzi}}, \ and\ \bibinfo {author}
  {\bibfnamefont {A.~G.}\ \bibnamefont {White}},\ }\href
  {https://dx.doi.org/10.1103/PhysRevLett.106.100401} {\bibfield  {journal}
  {\bibinfo  {journal} {Phys. Rev. Lett.}\ }\textbf {\bibinfo {volume} {106}},\
  \bibinfo {pages} {100401} (\bibinfo {year} {2011})}\BibitemShut {NoStop}%
\bibitem [{\citenamefont {Baldwin}\ \emph {et~al.}(2014)\citenamefont
  {Baldwin}, \citenamefont {Kalev},\ and\ \citenamefont
  {Deutsch}}]{Baldwin:2014aa}%
  \BibitemOpen
  \bibfield  {author} {\bibinfo {author} {\bibfnamefont {C.~H.}\ \bibnamefont
  {Baldwin}}, \bibinfo {author} {\bibfnamefont {A.}~\bibnamefont {Kalev}}, \
  and\ \bibinfo {author} {\bibfnamefont {I.~H.}\ \bibnamefont {Deutsch}},\
  }\href {https://dx.doi.org/10.1103/PhysRevA.90.012110} {\bibfield  {journal}
  {\bibinfo  {journal} {Phys. Rev. A}\ }\textbf {\bibinfo {volume} {90}},\
  \bibinfo {pages} {012110} (\bibinfo {year} {2014})}\BibitemShut {NoStop}%
\bibitem [{\citenamefont {Rodionov}\ \emph {et~al.}(2014)\citenamefont
  {Rodionov}, \citenamefont {Veitia}, \citenamefont {Barends}, \citenamefont
  {Kelly}, \citenamefont {Sank}, \citenamefont {Wenner}, \citenamefont
  {Martinis}, \citenamefont {Kosut},\ and\ \citenamefont
  {Korotkov}}]{Rodionov:2014aa}%
  \BibitemOpen
  \bibfield  {author} {\bibinfo {author} {\bibfnamefont {A.~V.}\ \bibnamefont
  {Rodionov}}, \bibinfo {author} {\bibfnamefont {A.}~\bibnamefont {Veitia}},
  \bibinfo {author} {\bibfnamefont {R.}~\bibnamefont {Barends}}, \bibinfo
  {author} {\bibfnamefont {J.}~\bibnamefont {Kelly}}, \bibinfo {author}
  {\bibfnamefont {D.}~\bibnamefont {Sank}}, \bibinfo {author} {\bibfnamefont
  {J.}~\bibnamefont {Wenner}}, \bibinfo {author} {\bibfnamefont {J.~M.}\
  \bibnamefont {Martinis}}, \bibinfo {author} {\bibfnamefont {R.~L.}\
  \bibnamefont {Kosut}}, \ and\ \bibinfo {author} {\bibfnamefont {A.~N.}\
  \bibnamefont {Korotkov}},\ }\href
  {https://dx.doi.org/10.1103/PhysRevB.90.144504} {\bibfield  {journal}
  {\bibinfo  {journal} {Phys. Rev. B}\ }\textbf {\bibinfo {volume} {90}},\
  \bibinfo {pages} {144504} (\bibinfo {year} {2014})}\BibitemShut {NoStop}%
\bibitem [{\citenamefont {Howland}\ and\ \citenamefont
  {Howell}(2013)}]{Howland:2013aa}%
  \BibitemOpen
  \bibfield  {author} {\bibinfo {author} {\bibfnamefont {G.~A.}\ \bibnamefont
  {Howland}}\ and\ \bibinfo {author} {\bibfnamefont {J.~C.}\ \bibnamefont
  {Howell}},\ }\href {\doibase 10.1103/PhysRevX.3.011013} {\bibfield  {journal}
  {\bibinfo  {journal} {Phys. Rev. X}\ }\textbf {\bibinfo {volume} {3}},\
  \bibinfo {pages} {011013} (\bibinfo {year} {2013})}\BibitemShut {NoStop}%
\bibitem [{\citenamefont {Howland}\ \emph {et~al.}(2014)\citenamefont
  {Howland}, \citenamefont {Schneeloch}, \citenamefont {Lum},\ and\
  \citenamefont {Howell}}]{Howland:2014aa}%
  \BibitemOpen
  \bibfield  {author} {\bibinfo {author} {\bibfnamefont {G.~A.}\ \bibnamefont
  {Howland}}, \bibinfo {author} {\bibfnamefont {J.}~\bibnamefont {Schneeloch}},
  \bibinfo {author} {\bibfnamefont {D.~J.}\ \bibnamefont {Lum}}, \ and\
  \bibinfo {author} {\bibfnamefont {J.~C.}\ \bibnamefont {Howell}},\ }\href
  {\doibase 10.1103/PhysRevLett.112.253602} {\bibfield  {journal} {\bibinfo
  {journal} {Phys. Rev. Lett.}\ }\textbf {\bibinfo {volume} {112}},\ \bibinfo
  {pages} {253602} (\bibinfo {year} {2014})}\BibitemShut {NoStop}%
\bibitem [{\citenamefont {Ahn}\ \emph {et~al.}(2019{\natexlab{a}})\citenamefont
  {Ahn}, \citenamefont {Teo}, \citenamefont {Jeong}, \citenamefont {Bouchard},
  \citenamefont {Hufnagel}, \citenamefont {Karimi}, \citenamefont {Koutn{\'y}},
  \citenamefont {{\v R}eh{\'a}{\v c}ek}, \citenamefont {Hradil}, \citenamefont
  {Leuchs},\ and\ \citenamefont {S{\'a}nchez-Soto}}]{Ahn:2019aa}%
  \BibitemOpen
  \bibfield  {author} {\bibinfo {author} {\bibfnamefont {D.}~\bibnamefont
  {Ahn}}, \bibinfo {author} {\bibfnamefont {Y.~S.}\ \bibnamefont {Teo}},
  \bibinfo {author} {\bibfnamefont {H.}~\bibnamefont {Jeong}}, \bibinfo
  {author} {\bibfnamefont {F.}~\bibnamefont {Bouchard}}, \bibinfo {author}
  {\bibfnamefont {F.}~\bibnamefont {Hufnagel}}, \bibinfo {author}
  {\bibfnamefont {E.}~\bibnamefont {Karimi}}, \bibinfo {author} {\bibfnamefont
  {D.}~\bibnamefont {Koutn{\'y}}}, \bibinfo {author} {\bibfnamefont
  {J.}~\bibnamefont {{\v R}eh{\'a}{\v c}ek}}, \bibinfo {author} {\bibfnamefont
  {Z.}~\bibnamefont {Hradil}}, \bibinfo {author} {\bibfnamefont
  {G.}~\bibnamefont {Leuchs}}, \ and\ \bibinfo {author} {\bibfnamefont {L.~L.}\
  \bibnamefont {S{\'a}nchez-Soto}},\ }\href
  {https://dx.doi.org/10.1103/PhysRevLett.122.100404} {\bibfield  {journal}
  {\bibinfo  {journal} {Phys. Rev. Lett.}\ }\textbf {\bibinfo {volume} {122}},\
  \bibinfo {pages} {100404} (\bibinfo {year} {2019}{\natexlab{a}})}\BibitemShut
  {NoStop}%
\bibitem [{\citenamefont {Ahn}\ \emph {et~al.}(2019{\natexlab{b}})\citenamefont
  {Ahn}, \citenamefont {Teo}, \citenamefont {Jeong}, \citenamefont
  {Koutn{\'y}}, \citenamefont {{\v R}eh{\'a}{\v c}ek}, \citenamefont {Hradil},
  \citenamefont {Leuchs},\ and\ \citenamefont {S{\'a}nchez-Soto}}]{Ahn:2019ns}%
  \BibitemOpen
  \bibfield  {author} {\bibinfo {author} {\bibfnamefont {D.}~\bibnamefont
  {Ahn}}, \bibinfo {author} {\bibfnamefont {Y.~S.}\ \bibnamefont {Teo}},
  \bibinfo {author} {\bibfnamefont {H.}~\bibnamefont {Jeong}}, \bibinfo
  {author} {\bibfnamefont {D.}~\bibnamefont {Koutn{\'y}}}, \bibinfo {author}
  {\bibfnamefont {J.}~\bibnamefont {{\v R}eh{\'a}{\v c}ek}}, \bibinfo {author}
  {\bibfnamefont {Z.}~\bibnamefont {Hradil}}, \bibinfo {author} {\bibfnamefont
  {G.}~\bibnamefont {Leuchs}}, \ and\ \bibinfo {author} {\bibfnamefont {L.~L.}\
  \bibnamefont {S{\'a}nchez-Soto}},\ }\href
  {https://dx.doi.org/10.1103/PhysRevA.100.012346} {\bibfield  {journal}
  {\bibinfo  {journal} {Phys. Rev. A}\ }\textbf {\bibinfo {volume} {100}},\
  \bibinfo {pages} {012346} (\bibinfo {year} {2019}{\natexlab{b}})}\BibitemShut
  {NoStop}%
\bibitem [{\citenamefont {Teo}\ \emph {et~al.}(2020)\citenamefont {Teo},
  \citenamefont {Struchalin}, \citenamefont {Kovlakov}, \citenamefont {Ahn},
  \citenamefont {Jeong}, \citenamefont {Straupe}, \citenamefont {Kulik},
  \citenamefont {Leuchs},\ and\ \citenamefont {S\'anchez-Soto}}]{Teo:2020cs}%
  \BibitemOpen
  \bibfield  {author} {\bibinfo {author} {\bibfnamefont {Y.~S.}\ \bibnamefont
  {Teo}}, \bibinfo {author} {\bibfnamefont {G.~I.}\ \bibnamefont {Struchalin}},
  \bibinfo {author} {\bibfnamefont {E.~V.}\ \bibnamefont {Kovlakov}}, \bibinfo
  {author} {\bibfnamefont {D.}~\bibnamefont {Ahn}}, \bibinfo {author}
  {\bibfnamefont {H.}~\bibnamefont {Jeong}}, \bibinfo {author} {\bibfnamefont
  {S.~S.}\ \bibnamefont {Straupe}}, \bibinfo {author} {\bibfnamefont {S.~P.}\
  \bibnamefont {Kulik}}, \bibinfo {author} {\bibfnamefont {G.}~\bibnamefont
  {Leuchs}}, \ and\ \bibinfo {author} {\bibfnamefont {L.~L.}\ \bibnamefont
  {S\'anchez-Soto}},\ }\href {\doibase 10.1103/PhysRevA.101.022334} {\bibfield
  {journal} {\bibinfo  {journal} {Phys. Rev. A}\ }\textbf {\bibinfo {volume}
  {101}},\ \bibinfo {pages} {022334} (\bibinfo {year} {2020})}\BibitemShut
  {NoStop}%
\bibitem [{\citenamefont {Kim}\ \emph {et~al.}(2020)\citenamefont {Kim},
  \citenamefont {Teo}, \citenamefont {Ahn}, \citenamefont {Im}, \citenamefont
  {Cho}, \citenamefont {Leuchs}, \citenamefont {S\'anchez-Soto}, \citenamefont
  {Jeong},\ and\ \citenamefont {Kim}}]{Kim:2020aa}%
  \BibitemOpen
  \bibfield  {author} {\bibinfo {author} {\bibfnamefont {Y.}~\bibnamefont
  {Kim}}, \bibinfo {author} {\bibfnamefont {Y.~S.}\ \bibnamefont {Teo}},
  \bibinfo {author} {\bibfnamefont {D.}~\bibnamefont {Ahn}}, \bibinfo {author}
  {\bibfnamefont {D.-G.}\ \bibnamefont {Im}}, \bibinfo {author} {\bibfnamefont
  {Y.-W.}\ \bibnamefont {Cho}}, \bibinfo {author} {\bibfnamefont
  {G.}~\bibnamefont {Leuchs}}, \bibinfo {author} {\bibfnamefont {L.~L.}\
  \bibnamefont {S\'anchez-Soto}}, \bibinfo {author} {\bibfnamefont
  {H.}~\bibnamefont {Jeong}}, \ and\ \bibinfo {author} {\bibfnamefont {Y.-H.}\
  \bibnamefont {Kim}},\ }\href {\doibase 10.1103/PhysRevLett.124.210401}
  {\bibfield  {journal} {\bibinfo  {journal} {Phys. Rev. Lett.}\ }\textbf
  {\bibinfo {volume} {124}},\ \bibinfo {pages} {210401} (\bibinfo {year}
  {2020})}\BibitemShut {NoStop}%
\bibitem [{\citenamefont {Eldar}\ \emph {et~al.}(2012)\citenamefont {Eldar},
  \citenamefont {Needell},\ and\ \citenamefont {Plan}}]{ELDAR:2012aa}%
  \BibitemOpen
  \bibfield  {author} {\bibinfo {author} {\bibfnamefont {Y.}~\bibnamefont
  {Eldar}}, \bibinfo {author} {\bibfnamefont {D.}~\bibnamefont {Needell}}, \
  and\ \bibinfo {author} {\bibfnamefont {Y.}~\bibnamefont {Plan}},\ }\href
  {\doibase https://doi.org/10.1016/j.acha.2012.04.002} {\bibfield  {journal}
  {\bibinfo  {journal} {Appl. Comput. Harmon. Anal.}\ }\textbf {\bibinfo
  {volume} {33}},\ \bibinfo {pages} {309 } (\bibinfo {year}
  {2012})}\BibitemShut {NoStop}%
\bibitem [{\citenamefont {Bandeira}\ \emph {et~al.}(2014)\citenamefont
  {Bandeira}, \citenamefont {Cahill}, \citenamefont {Mixon},\ and\
  \citenamefont {Nelson}}]{BANDEIRA:2014aa}%
  \BibitemOpen
  \bibfield  {author} {\bibinfo {author} {\bibfnamefont {A.~S.}\ \bibnamefont
  {Bandeira}}, \bibinfo {author} {\bibfnamefont {J.}~\bibnamefont {Cahill}},
  \bibinfo {author} {\bibfnamefont {D.~G.}\ \bibnamefont {Mixon}}, \ and\
  \bibinfo {author} {\bibfnamefont {A.~A.}\ \bibnamefont {Nelson}},\ }\href
  {\doibase https://doi.org/10.1016/j.acha.2013.10.002} {\bibfield  {journal}
  {\bibinfo  {journal} {Appl. Comput. Harmon. Anal.}\ }\textbf {\bibinfo
  {volume} {37}},\ \bibinfo {pages} {106 } (\bibinfo {year}
  {2014})}\BibitemShut {NoStop}%
\bibitem [{\citenamefont {Bodmann}\ and\ \citenamefont
  {Hammen}(2015)}]{Bodmann:2015aa}%
  \BibitemOpen
  \bibfield  {author} {\bibinfo {author} {\bibfnamefont {B.~G.}\ \bibnamefont
  {Bodmann}}\ and\ \bibinfo {author} {\bibfnamefont {N.}~\bibnamefont
  {Hammen}},\ }\href {\doibase https://doi.org/10.1007/s10444-014-9359-y}
  {\bibfield  {journal} {\bibinfo  {journal} {Adv. Comput. Math.}\ }\textbf
  {\bibinfo {volume} {41}},\ \bibinfo {pages} {317 } (\bibinfo {year}
  {2015})}\BibitemShut {NoStop}%
\bibitem [{\citenamefont {Xu}(2018)}]{Xu:2018aa}%
  \BibitemOpen
  \bibfield  {author} {\bibinfo {author} {\bibfnamefont {Z.}~\bibnamefont
  {Xu}},\ }\href {https://doi.org/10.1016/j.acha.2017.01.005} {\bibfield
  {journal} {\bibinfo  {journal} {Appl. Comput. Harmon. Anal.}\ }\textbf
  {\bibinfo {volume} {44}},\ \bibinfo {pages} {497 } (\bibinfo {year}
  {2018})}\BibitemShut {NoStop}%
\bibitem [{\citenamefont {Palsson}\ \emph {et~al.}(2017)\citenamefont
  {Palsson}, \citenamefont {Gu}, \citenamefont {Ho}, \citenamefont {Wiseman},\
  and\ \citenamefont {Pryde}}]{Palsson}%
  \BibitemOpen
  \bibfield  {author} {\bibinfo {author} {\bibfnamefont {M.~S.}\ \bibnamefont
  {Palsson}}, \bibinfo {author} {\bibfnamefont {M.}~\bibnamefont {Gu}},
  \bibinfo {author} {\bibfnamefont {J.}~\bibnamefont {Ho}}, \bibinfo {author}
  {\bibfnamefont {H.~M.}\ \bibnamefont {Wiseman}}, \ and\ \bibinfo {author}
  {\bibfnamefont {G.~J.}\ \bibnamefont {Pryde}},\ }\href {\doibase
  10.1126/sciadv.1601302} {\bibfield  {journal} {\bibinfo  {journal} {Sci.
  Adv.}\ }\textbf {\bibinfo {volume} {3}},\ \bibinfo {pages} {e1601302}
  (\bibinfo {year} {2017})}\BibitemShut {NoStop}%
\bibitem [{\citenamefont {{\v R}eh{\'a}{\v c}ek}\ \emph
  {et~al.}(2007)\citenamefont {{\v R}eh{\'a}{\v c}ek}, \citenamefont {Hradil},
  \citenamefont {Knill},\ and\ \citenamefont {Lvovsky}}]{Rehacek:2007ml}%
  \BibitemOpen
  \bibfield  {author} {\bibinfo {author} {\bibfnamefont {J.}~\bibnamefont {{\v
  R}eh{\'a}{\v c}ek}}, \bibinfo {author} {\bibfnamefont {Z.}~\bibnamefont
  {Hradil}}, \bibinfo {author} {\bibfnamefont {E.}~\bibnamefont {Knill}}, \
  and\ \bibinfo {author} {\bibfnamefont {A.~I.}\ \bibnamefont {Lvovsky}},\
  }\href {https://dx.doi.org/10.1103/PhysRevA.75.042108} {\bibfield  {journal}
  {\bibinfo  {journal} {Phys. Rev. A}\ }\textbf {\bibinfo {volume} {75}},\
  \bibinfo {pages} {042108} (\bibinfo {year} {2007})}\BibitemShut {NoStop}%
\bibitem [{\citenamefont {Teo}\ \emph {et~al.}(2011)\citenamefont {Teo},
  \citenamefont {Zhu}, \citenamefont {Englert}, \citenamefont {{\v R}eh{\'a}{\v
  c}ek},\ and\ \citenamefont {Hradil}}]{Teo:2011me}%
  \BibitemOpen
  \bibfield  {author} {\bibinfo {author} {\bibfnamefont {Y.~S.}\ \bibnamefont
  {Teo}}, \bibinfo {author} {\bibfnamefont {H.}~\bibnamefont {Zhu}}, \bibinfo
  {author} {\bibfnamefont {B.-G.}\ \bibnamefont {Englert}}, \bibinfo {author}
  {\bibfnamefont {J.}~\bibnamefont {{\v R}eh{\'a}{\v c}ek}}, \ and\ \bibinfo
  {author} {\bibfnamefont {Z.}~\bibnamefont {Hradil}},\ }\href
  {https://dx.doi.org/10.1103/PhysRevLett.107.020404} {\bibfield  {journal}
  {\bibinfo  {journal} {Phys. Rev. Lett.}\ }\textbf {\bibinfo {volume} {107}},\
  \bibinfo {pages} {020404} (\bibinfo {year} {2011})}\BibitemShut {NoStop}%
\bibitem [{\citenamefont {Shang}\ \emph {et~al.}(2017)\citenamefont {Shang},
  \citenamefont {Zhang},\ and\ \citenamefont {Ng}}]{Shang:2017sf}%
  \BibitemOpen
  \bibfield  {author} {\bibinfo {author} {\bibfnamefont {J.}~\bibnamefont
  {Shang}}, \bibinfo {author} {\bibfnamefont {Z.}~\bibnamefont {Zhang}}, \ and\
  \bibinfo {author} {\bibfnamefont {H.~K.}\ \bibnamefont {Ng}},\ }\href
  {https://dx.doi.org/10.1103/PhysRevA.95.062336} {\bibfield  {journal}
  {\bibinfo  {journal} {Phys. Rev. A}\ }\textbf {\bibinfo {volume} {95}},\
  \bibinfo {pages} {062336} (\bibinfo {year} {2017})}\BibitemShut {NoStop}%
\bibitem [{\citenamefont {Vandenberghe}\ and\ \citenamefont
  {Boyd}(1996)}]{Vandenberghe:1996ca}%
  \BibitemOpen
  \bibfield  {author} {\bibinfo {author} {\bibfnamefont {L.}~\bibnamefont
  {Vandenberghe}}\ and\ \bibinfo {author} {\bibfnamefont {S.}~\bibnamefont
  {Boyd}},\ }\href {https://doi.org/10.1137/1038003} {\bibfield  {journal}
  {\bibinfo  {journal} {SIAM Rev.}\ }\textbf {\bibinfo {volume} {38}},\
  \bibinfo {pages} {49} (\bibinfo {year} {1996})}\BibitemShut {NoStop}%
\bibitem [{\citenamefont {Jackson}\ and\ \citenamefont {van
  Enk}(2015)}]{Jackson:2015aa}%
  \BibitemOpen
  \bibfield  {author} {\bibinfo {author} {\bibfnamefont {C.}~\bibnamefont
  {Jackson}}\ and\ \bibinfo {author} {\bibfnamefont {S.~J.}\ \bibnamefont {van
  Enk}},\ }\href {\doibase 10.1103/PhysRevA.92.042312} {\bibfield  {journal}
  {\bibinfo  {journal} {Phys. Rev. A}\ }\textbf {\bibinfo {volume} {92}},\
  \bibinfo {pages} {042312} (\bibinfo {year} {2015})}\BibitemShut {NoStop}%
\bibitem [{\citenamefont {Ballance}\ \emph {et~al.}(2016)\citenamefont
  {Ballance}, \citenamefont {Harty}, \citenamefont {Linke}, \citenamefont
  {Sepiol},\ and\ \citenamefont {Lucas}}]{Ballance:2016aa}%
  \BibitemOpen
  \bibfield  {author} {\bibinfo {author} {\bibfnamefont {C.~J.}\ \bibnamefont
  {Ballance}}, \bibinfo {author} {\bibfnamefont {T.~P.}\ \bibnamefont {Harty}},
  \bibinfo {author} {\bibfnamefont {N.~M.}\ \bibnamefont {Linke}}, \bibinfo
  {author} {\bibfnamefont {M.~A.}\ \bibnamefont {Sepiol}}, \ and\ \bibinfo
  {author} {\bibfnamefont {D.~M.}\ \bibnamefont {Lucas}},\ }\href {\doibase
  10.1103/PhysRevLett.117.060504} {\bibfield  {journal} {\bibinfo  {journal}
  {Phys. Rev. Lett.}\ }\textbf {\bibinfo {volume} {117}},\ \bibinfo {pages}
  {060504} (\bibinfo {year} {2016})}\BibitemShut {NoStop}%
\bibitem [{\citenamefont {Erhard}\ \emph {et~al.}(2019)\citenamefont {Erhard},
  \citenamefont {Wallman}, \citenamefont {Postler}, \citenamefont {Meth},
  \citenamefont {Stricker}, \citenamefont {Martinez}, \citenamefont
  {Schindler}, \citenamefont {Monz}, \citenamefont {J.},\ and\ \citenamefont
  {Blatt}}]{Erhard:2019aa}%
  \BibitemOpen
  \bibfield  {author} {\bibinfo {author} {\bibfnamefont {A.}~\bibnamefont
  {Erhard}}, \bibinfo {author} {\bibfnamefont {J.~J.}\ \bibnamefont {Wallman}},
  \bibinfo {author} {\bibfnamefont {L.}~\bibnamefont {Postler}}, \bibinfo
  {author} {\bibfnamefont {M.}~\bibnamefont {Meth}}, \bibinfo {author}
  {\bibfnamefont {R.}~\bibnamefont {Stricker}}, \bibinfo {author}
  {\bibfnamefont {E.~A.}\ \bibnamefont {Martinez}}, \bibinfo {author}
  {\bibfnamefont {P.}~\bibnamefont {Schindler}}, \bibinfo {author}
  {\bibfnamefont {T.}~\bibnamefont {Monz}}, \bibinfo {author} {\bibfnamefont
  {E.}~\bibnamefont {J.}}, \ and\ \bibinfo {author} {\bibfnamefont
  {R.}~\bibnamefont {Blatt}},\ }\href
  {https://doi.org/10.1038/s41467-019-13068-7} {\bibfield  {journal} {\bibinfo
  {journal} {Nat. Commun.}\ }\textbf {\bibinfo {volume} {10}},\ \bibinfo
  {pages} {5347} (\bibinfo {year} {2019})}\BibitemShut {NoStop}%
\bibitem [{\citenamefont {Hausladen}\ and\ \citenamefont
  {Wootters}(1994)}]{Hausladen:1994aa}%
  \BibitemOpen
  \bibfield  {author} {\bibinfo {author} {\bibfnamefont {P.}~\bibnamefont
  {Hausladen}}\ and\ \bibinfo {author} {\bibfnamefont {W.~K.}\ \bibnamefont
  {Wootters}},\ }\href {\doibase 10.1080/09500349414552221} {\bibfield
  {journal} {\bibinfo  {journal} {J. Mod. Opt.}\ }\textbf {\bibinfo {volume}
  {41}},\ \bibinfo {pages} {2385} (\bibinfo {year} {1994})}\BibitemShut
  {NoStop}%
\bibitem [{\citenamefont {{Eldar}}\ and\ \citenamefont
  {{Forney}}(2001)}]{Eldar:2001aa}%
  \BibitemOpen
  \bibfield  {author} {\bibinfo {author} {\bibfnamefont {Y.~C.}\ \bibnamefont
  {{Eldar}}}\ and\ \bibinfo {author} {\bibfnamefont {G.~D.}\ \bibnamefont
  {{Forney}}},\ }\href {https://doi.org/10.1109/18.915636} {\bibfield
  {journal} {\bibinfo  {journal} {IEEE Trans. Inf. Theory}\ }\textbf {\bibinfo
  {volume} {47}},\ \bibinfo {pages} {858 } (\bibinfo {year}
  {2001})}\BibitemShut {NoStop}%
\bibitem [{\citenamefont {Dalla~Pozza}\ and\ \citenamefont
  {Pierobon}(2015)}]{Pozza:2015aa}%
  \BibitemOpen
  \bibfield  {author} {\bibinfo {author} {\bibfnamefont {N.}~\bibnamefont
  {Dalla~Pozza}}\ and\ \bibinfo {author} {\bibfnamefont {G.}~\bibnamefont
  {Pierobon}},\ }\href {\doibase 10.1103/PhysRevA.91.042334} {\bibfield
  {journal} {\bibinfo  {journal} {Phys. Rev. A}\ }\textbf {\bibinfo {volume}
  {91}},\ \bibinfo {pages} {042334} (\bibinfo {year} {2015})}\BibitemShut
  {NoStop}%
\bibitem [{\citenamefont {Heinosaari}\ \emph {et~al.}(2020)\citenamefont
  {Heinosaari}, \citenamefont {Jivulescu},\ and\ \citenamefont
  {Nechita}}]{Heinosaari:2020aa}%
  \BibitemOpen
  \bibfield  {author} {\bibinfo {author} {\bibfnamefont {T.}~\bibnamefont
  {Heinosaari}}, \bibinfo {author} {\bibfnamefont {M.~A.}\ \bibnamefont
  {Jivulescu}}, \ and\ \bibinfo {author} {\bibfnamefont {I.}~\bibnamefont
  {Nechita}},\ }\href {\doibase 10.1063/1.5131028} {\bibfield  {journal}
  {\bibinfo  {journal} {J. Math. Phys.}\ }\textbf {\bibinfo {volume} {61}},\
  \bibinfo {pages} {042202} (\bibinfo {year} {2020})}\BibitemShut {NoStop}%
\bibitem [{\citenamefont {Langford}\ \emph {et~al.}(2005)\citenamefont
  {Langford}, \citenamefont {Weinhold}, \citenamefont {Prevedel}, \citenamefont
  {Resch}, \citenamefont {Gilchrist}, \citenamefont {O'Brien}, \citenamefont
  {Pryde},\ and\ \citenamefont {White}}]{nathan}%
  \BibitemOpen
  \bibfield  {author} {\bibinfo {author} {\bibfnamefont {N.~K.}\ \bibnamefont
  {Langford}}, \bibinfo {author} {\bibfnamefont {T.~J.}\ \bibnamefont
  {Weinhold}}, \bibinfo {author} {\bibfnamefont {R.}~\bibnamefont {Prevedel}},
  \bibinfo {author} {\bibfnamefont {K.~J.}\ \bibnamefont {Resch}}, \bibinfo
  {author} {\bibfnamefont {A.}~\bibnamefont {Gilchrist}}, \bibinfo {author}
  {\bibfnamefont {J.~L.}\ \bibnamefont {O'Brien}}, \bibinfo {author}
  {\bibfnamefont {G.~J.}\ \bibnamefont {Pryde}}, \ and\ \bibinfo {author}
  {\bibfnamefont {A.~G.}\ \bibnamefont {White}},\ }\href {\doibase
  10.1103/PhysRevLett.95.210504} {\bibfield  {journal} {\bibinfo  {journal}
  {Phys. Rev. Lett.}\ }\textbf {\bibinfo {volume} {95}},\ \bibinfo {pages}
  {210504} (\bibinfo {year} {2005})}\BibitemShut {NoStop}%
\bibitem [{\citenamefont {Kiesel}\ \emph {et~al.}(2005)\citenamefont {Kiesel},
  \citenamefont {Schmid}, \citenamefont {Weber}, \citenamefont {Ursin},\ and\
  \citenamefont {Weinfurter}}]{kiesel}%
  \BibitemOpen
  \bibfield  {author} {\bibinfo {author} {\bibfnamefont {N.}~\bibnamefont
  {Kiesel}}, \bibinfo {author} {\bibfnamefont {C.}~\bibnamefont {Schmid}},
  \bibinfo {author} {\bibfnamefont {U.}~\bibnamefont {Weber}}, \bibinfo
  {author} {\bibfnamefont {R.}~\bibnamefont {Ursin}}, \ and\ \bibinfo {author}
  {\bibfnamefont {H.}~\bibnamefont {Weinfurter}},\ }\href {\doibase
  10.1103/PhysRevLett.95.210505} {\bibfield  {journal} {\bibinfo  {journal}
  {Phys. Rev. Lett.}\ }\textbf {\bibinfo {volume} {95}},\ \bibinfo {pages}
  {210505} (\bibinfo {year} {2005})}\BibitemShut {NoStop}%
\bibitem [{\citenamefont {Okamoto}\ \emph {et~al.}(2005)\citenamefont
  {Okamoto}, \citenamefont {Hofmann}, \citenamefont {Takeuchi},\ and\
  \citenamefont {Sasaki}}]{okamoto}%
  \BibitemOpen
  \bibfield  {author} {\bibinfo {author} {\bibfnamefont {R.}~\bibnamefont
  {Okamoto}}, \bibinfo {author} {\bibfnamefont {H.~F.}\ \bibnamefont
  {Hofmann}}, \bibinfo {author} {\bibfnamefont {S.}~\bibnamefont {Takeuchi}}, \
  and\ \bibinfo {author} {\bibfnamefont {K.}~\bibnamefont {Sasaki}},\ }\href
  {\doibase 10.1103/PhysRevLett.95.210506} {\bibfield  {journal} {\bibinfo
  {journal} {Phys. Rev. Lett.}\ }\textbf {\bibinfo {volume} {95}},\ \bibinfo
  {pages} {210506} (\bibinfo {year} {2005})}\BibitemShut {NoStop}%
\bibitem [{\citenamefont {Roccia}\ \emph {et~al.}(2017)\citenamefont {Roccia},
  \citenamefont {Gianani}, \citenamefont {Mancino}, \citenamefont {Sbroscia},
  \citenamefont {Somma}, \citenamefont {Genoni},\ and\ \citenamefont
  {Barbieri}}]{Roccia:2017aa}%
  \BibitemOpen
  \bibfield  {author} {\bibinfo {author} {\bibfnamefont {E.}~\bibnamefont
  {Roccia}}, \bibinfo {author} {\bibfnamefont {I.}~\bibnamefont {Gianani}},
  \bibinfo {author} {\bibfnamefont {L.}~\bibnamefont {Mancino}}, \bibinfo
  {author} {\bibfnamefont {M.}~\bibnamefont {Sbroscia}}, \bibinfo {author}
  {\bibfnamefont {F.}~\bibnamefont {Somma}}, \bibinfo {author} {\bibfnamefont
  {M.~G.}\ \bibnamefont {Genoni}}, \ and\ \bibinfo {author} {\bibfnamefont
  {M.}~\bibnamefont {Barbieri}},\ }\href {\doibase 10.1088/2058-9565/aa9212}
  {\bibfield  {journal} {\bibinfo  {journal} {Quantum Sci. Technol.}\ }\textbf
  {\bibinfo {volume} {3}},\ \bibinfo {pages} {01LT01} (\bibinfo {year}
  {2017})}\BibitemShut {NoStop}%
\bibitem [{\citenamefont {Chuang}\ and\ \citenamefont
  {Nielsen}(2000)}]{Chuang:2000fk}%
  \BibitemOpen
  \bibfield  {author} {\bibinfo {author} {\bibfnamefont {I.}~\bibnamefont
  {Chuang}}\ and\ \bibinfo {author} {\bibfnamefont {M.}~\bibnamefont
  {Nielsen}},\ }\href@noop {} {\emph {\bibinfo {title} {Quantum Computation and
  Quantum Information}}}\ (\bibinfo  {publisher} {Cambridge University Press},\
  \bibinfo {address} {Cambridge},\ \bibinfo {year} {2000})\BibitemShut
  {NoStop}%
\bibitem [{\citenamefont {Hayashi}(1998)}]{Hayashi:1998aa}%
  \BibitemOpen
  \bibfield  {author} {\bibinfo {author} {\bibfnamefont {M.}~\bibnamefont
  {Hayashi}},\ }\href {\doibase 10.1088/0305-4470/31/20/006} {\bibfield
  {journal} {\bibinfo  {journal} {J. Phys. A: Math and Gen.}\ }\textbf
  {\bibinfo {volume} {31}},\ \bibinfo {pages} {4633} (\bibinfo {year}
  {1998})}\BibitemShut {NoStop}%
\bibitem [{\citenamefont {D'Ariano}\ and\ \citenamefont
  {Perinotti}(2007)}]{D'Ariano:2007aa}%
  \BibitemOpen
  \bibfield  {author} {\bibinfo {author} {\bibfnamefont {G.~M.}\ \bibnamefont
  {D'Ariano}}\ and\ \bibinfo {author} {\bibfnamefont {P.}~\bibnamefont
  {Perinotti}},\ }\href {\doibase 10.1103/PhysRevLett.98.020403} {\bibfield
  {journal} {\bibinfo  {journal} {Phys. Rev. Lett.}\ }\textbf {\bibinfo
  {volume} {98}},\ \bibinfo {pages} {020403} (\bibinfo {year}
  {2007})}\BibitemShut {NoStop}%
\bibitem [{\citenamefont {Bisio}\ \emph {et~al.}(2009)\citenamefont {Bisio},
  \citenamefont {Chiribella}, \citenamefont {D'Ariano}, \citenamefont
  {Facchini},\ and\ \citenamefont {Perinotti}}]{Bisio:2009aa}%
  \BibitemOpen
  \bibfield  {author} {\bibinfo {author} {\bibfnamefont {A.}~\bibnamefont
  {Bisio}}, \bibinfo {author} {\bibfnamefont {G.}~\bibnamefont {Chiribella}},
  \bibinfo {author} {\bibfnamefont {G.~M.}\ \bibnamefont {D'Ariano}}, \bibinfo
  {author} {\bibfnamefont {S.}~\bibnamefont {Facchini}}, \ and\ \bibinfo
  {author} {\bibfnamefont {P.}~\bibnamefont {Perinotti}},\ }\href {\doibase
  10.1103/PhysRevLett.102.010404} {\bibfield  {journal} {\bibinfo  {journal}
  {Phys. Rev. Lett.}\ }\textbf {\bibinfo {volume} {102}},\ \bibinfo {pages}
  {010404} (\bibinfo {year} {2009})}\BibitemShut {NoStop}%
\bibitem [{\citenamefont {{Bisio}}\ \emph {et~al.}(2009)\citenamefont
  {{Bisio}}, \citenamefont {{Chiribella}}, \citenamefont {{D'Ariano}},
  \citenamefont {{Facchini}},\ and\ \citenamefont
  {{Perinotti}}}]{Bisio:2009qt}%
  \BibitemOpen
  \bibfield  {author} {\bibinfo {author} {\bibfnamefont {A.}~\bibnamefont
  {{Bisio}}}, \bibinfo {author} {\bibfnamefont {G.}~\bibnamefont
  {{Chiribella}}}, \bibinfo {author} {\bibfnamefont {G.~M.}\ \bibnamefont
  {{D'Ariano}}}, \bibinfo {author} {\bibfnamefont {S.}~\bibnamefont
  {{Facchini}}}, \ and\ \bibinfo {author} {\bibfnamefont {P.}~\bibnamefont
  {{Perinotti}}},\ }\href {https://doi.org/10.1109/JSTQE.2009.2029243}
  {\bibfield  {journal} {\bibinfo  {journal} {IEEE Journal of Selected Topics
  in Quantum Electronics}\ }\textbf {\bibinfo {volume} {15}},\ \bibinfo {pages}
  {1646} (\bibinfo {year} {2009})}\BibitemShut {NoStop}%
\bibitem [{\citenamefont {Teo}(2015)}]{Teo:2015qs}%
  \BibitemOpen
  \bibfield  {author} {\bibinfo {author} {\bibfnamefont {Y.~S.}\ \bibnamefont
  {Teo}},\ }\href@noop {} {\emph {\bibinfo {title} {Introduction to
  {Q}uantum-{S}tate {E}stimation}}}\ (\bibinfo  {publisher} {World Scientific
  Publishing Co.},\ \bibinfo {address} {Singapore},\ \bibinfo {year}
  {2015})\BibitemShut {NoStop}%
\bibitem [{\citenamefont {De~Santis}\ \emph {et~al.}(2019)\citenamefont
  {De~Santis}, \citenamefont {Coppola}, \citenamefont {Ant\'on}, \citenamefont
  {Somaschi}, \citenamefont {G\'omez}, \citenamefont {Lema\^{\i}tre},
  \citenamefont {Sagnes}, \citenamefont {Lanco}, \citenamefont {Loredo},
  \citenamefont {Krebs},\ and\ \citenamefont {Senellart}}]{Santis:2019aa}%
  \BibitemOpen
  \bibfield  {author} {\bibinfo {author} {\bibfnamefont {L.}~\bibnamefont
  {De~Santis}}, \bibinfo {author} {\bibfnamefont {G.}~\bibnamefont {Coppola}},
  \bibinfo {author} {\bibfnamefont {C.}~\bibnamefont {Ant\'on}}, \bibinfo
  {author} {\bibfnamefont {N.}~\bibnamefont {Somaschi}}, \bibinfo {author}
  {\bibfnamefont {C.}~\bibnamefont {G\'omez}}, \bibinfo {author} {\bibfnamefont
  {A.}~\bibnamefont {Lema\^{\i}tre}}, \bibinfo {author} {\bibfnamefont
  {I.}~\bibnamefont {Sagnes}}, \bibinfo {author} {\bibfnamefont
  {L.}~\bibnamefont {Lanco}}, \bibinfo {author} {\bibfnamefont {J.~C.}\
  \bibnamefont {Loredo}}, \bibinfo {author} {\bibfnamefont {O.}~\bibnamefont
  {Krebs}}, \ and\ \bibinfo {author} {\bibfnamefont {P.}~\bibnamefont
  {Senellart}},\ }\href {\doibase 10.1103/PhysRevA.99.022312} {\bibfield
  {journal} {\bibinfo  {journal} {Phys. Rev. A}\ }\textbf {\bibinfo {volume}
  {99}},\ \bibinfo {pages} {022312} (\bibinfo {year} {2019})}\BibitemShut
  {NoStop}%
\end{thebibliography}
\end{document}